\documentclass[aps,twocolumn,prc,amssymb,amsmath,showpacs]{revtex4}

\usepackage[10pt]{type1ec}  % use only 10pt fonts
\usepackage[T1]{fontenc}
\usepackage{CJKutf8}
\usepackage[overlap, CJK]{ruby}
\usepackage{CJKulem}
\newenvironment{SChinese}{%
  \CJKfamily{gbsn}%
  \CJKtilde
  \CJKnospace}{}
\usepackage{graphicx,stackengine,xcolor}
\usepackage{multirow}
\usepackage[breaklinks,colorlinks,urlcolor=blue,citecolor=blue]{hyperref}
\usepackage{mathrsfs}
\usepackage{soul}
\usepackage{color}

\usepackage[utf8]{inputenc}
\usepackage{bm}                      % bold math
\usepackage{dcolumn}                 % Align table columns on decimal point

\newcommand{\al}{\alpha}
\newcommand{\be}{\beta}
\newcommand{\ga}{\gamma}
\newcommand{\Ga}{\Gamma}

\newcommand{\De}{\Delta}
\newcommand{\ep}{\varepsilon}
\newcommand{\eps}{\epsilon}
\newcommand{\Msolar}{{\rm M}_{\odot}}
\newcommand{\ts}{T_s}
\newcommand{\tb}{T_b}
\newcommand{\ti}{T_i}
\newcommand{\tc}{T_c}
\newcommand{\nb}{n_{\rm B}}
\newcommand{\sft}{^{3}P_2}
\newcommand{\sfs}{^{1}S_0}

\newcommand{\beq}{\begin{equation}}
\newcommand{\eeq}{\end{equation}}
\newcommand{\ba}{\begin{array}}
\newcommand{\ea}{\end{array}}
\newcommand{\bea}{\begin{eqnarray}}
\newcommand{\eea}{\end{eqnarray}}
\newcommand{\bc}{\begin{center}}
\newcommand{\ec}{\end{center}}

\newcommand{\dsp}{\displaystyle}
\newcommand\eqn[1]{(\ref{#1})}      % parentheses around the LaTex "ref" macro
\newcommand\Eqn[1]{Eq.~(\ref{#1})}  % includes ``Eq.'' in front
\newcommand{\e}{{\rm e}}   % 2.718281828
\newcommand{\nn}{\nonumber \\}
\newcommand{\ee}[1]{\times 10^{#1}}

\begin{document}

\title{Cooling of neutron stars in soft X-ray transients} 

\author{Sophia Han (\begin{CJK}{UTF8}{}\begin{SChinese}韩 君\end{SChinese}\end{CJK})$^{1}$} 
\author{Andrew W. Steiner$^{1,2}$} 

\affiliation{$^{1}$Department of Physics and Astronomy, University of
  Tennessee, Knoxville, TN 37996, USA}
\affiliation{$^{2}$Physics Division, Oak Ridge National Laboratory, Oak
  Ridge, TN 37831, USA}

\date{01 August 2017} % hardwire date so arxiv can't change it

\begin{abstract}
Thermal states of neutron stars in soft X-ray transients (SXRTs) are
thought to be determined by ``deep crustal heating'' in the accreted
matter that drives the quiescent luminosity and cooling via emission
of photons and neutrinos from the interior. In this study, we assume a
global thermal steady-state of the transient system and calculate the
heating curves (quiescent surface luminosity as a function of mean
accretion rate) predicted from theoretical models, taking into account
variations in the equations of state, superfluidity gaps, thickness of
the light element layer and a phenomenological description of the
direct Urca threshold. We further provide a statistical analysis on the uncertainties in these parameters, and compare the overall results
with observations of several SXRTs, in particular the two sources
containing the coldest (SAX J1808.4-3658) and the hottest (Aql X-1)
neutron stars. Interpretation of the observational data indicates that
the direct Urca process is required for the most massive stars and also 
suggests small superfluid gaps.
\end{abstract}


\maketitle
%===============================================================================
\section{Introduction}
\label{sec:intro}

One of the outstanding questions in strong interaction physics is the nature of matter that is denser than the centers of atomic nuclei. Neutron stars are an excellent laboratory of cold and dense matter, and thus it is of interest to attempt to probe the nature of dense matter from neutron star observations. At nuclear densities, the ground state of matter consists of neutrons and protons, but at higher densities several different possibilities emerge: hyperons or Bose condensates may be present or hadrons may become deconfined leaving behind quark matter. This uncertainty regarding the nature of dense matter creates an overwhelmingly large space of possible theoretical models of neutron star structure and evolution. The large uncertainty can be tamed by assuming a ``minimal model'' of dense matter where neutron stars are assumed to contain only neutrons, protons, and electrons.

As originally formulated in Ref.~\cite{Page04}, the minimal model
additionally assumes no cooling from the so-called direct Urca
process~\cite{boguta81,Lattimer:1991ib}, and thus assuming that the 
number of protons in dense matter is not large enough to allow the direct 
Urca neutrino emission to cool the star. Historically, this assumption was 
sensible because, previous to a decade ago, most neutron star mass 
measurements were near $1.4\,\Msolar$ and therefore neutron star 
densities need not be large enough to allow the direct Urca process to 
occur. This has changed dramatically with the discovery of two neutron 
stars with masses near $2\,\Msolar$~\cite{Demorest10,Antoniadis13}.

Beyond high-mass neuron stars, the most significant challenge to the
minimal model arose in 2007 when it was confirmed that some accreting
neutron stars are exceptionally cold, in spite of periodic, appreciable accretion from a main-sequence companion~\cite{Heinke:2006ie}. This work found that, during quiescence when accretion and its concomitant X-ray emission have stopped, SAX J1808.4-3658 is much colder than expected from the minimal model. In contrast, the cooling of isolated neutron stars with no companion is easier to explain in the minimal model, so long as a contribution from nucleon superfluidity was included~\cite{Page04,Page:2009fu}.

One possible explanation for these results is that accreting neutron
stars are more massive, have cores which are more proton-rich, and
thus are colder because they cool through direct Urca. Isolated neutron stars, on the other hand, have not accreted and are less massive and their cores follow the minimal model, or else their cores are too cold to be seen in the X-ray band~\cite{Kaplan:2004tm,Kaplan:2006mb}. In this paper, we explore this possibility and determine the most likely values of the pressure of dense matter, the size of the nucleon superfluid critical temperatures in the core, the mass of SAX J1808.4-3658, the threshold density for the
direct Urca process, and other quantities of interest.

\section{method and formalism}
\label{sec:theory}

\subsection{Thermal states of SXRTs}

The SXRTs typically undergo quiescence intervals much shorter than
neutron star thermal relaxation timescale $\lesssim10^4$ yrs
\cite{lattimer1994rapid,Colpi:2000jc,Wijnands:2012tf}, therefore it 
is often plausible to ignore thermal variations in the core and treat their internal temperature as independent of the transient accretion (cf. studies of crustal cooling, e.g.~\cite{Cackett:2006qk}). We calculate the thermal structure of transiently accreting neutron stars in a quiescent state, which is mainly governed by the deep crustal heating~\cite{Haensel90,Brown:1998qv} due to nuclear transformations in the accreted crust and cooling processes of neutrino and photon emission at the neutron star interior/surface. The heat balance equation reads
\beq
L_{\rm dh}^{\infty} (\dot{M})=
L_{\ga}^{\infty}(\ts)+L_{\nu}^{\infty}(\ti)
\label{eqn:hc}
\eeq
where $L_{\rm dh}^{\infty}$ is the red-shifted power of the deep crustal heating, $L_{\ga}^{\infty}$ is the effective surface luminosity as detected by a distant observer, and $L_{\nu}^{\infty}$ is the red-shifted neutrino luminosity determined by the internal structure and composition of the star;
\bea
L_{\ga}^{\infty}&=& \dsp{4\pi\sigma\ts^{4}R^{2}(1-2\,GM/(c^2
    R))} \label{eqn:Lum-ph} \\
L_{\nu}^{\infty}&=& 4\pi\int_{0}^{R}\frac{\mathrm{d} {r} \,r^{2} \eps_{\nu} \e^{2\Phi(r)}}{\sqrt{1-2\,Gm(r)/(c^2\,r)}}
\label{eqn:Lum-nu}
\eea
$m(r)$ is the enclosed gravitational mass and $\Phi(r)$ is the metric
function.

The mean accretion rate $\dot{M}\equiv t_a \dot{M_a}/(t_a+t_q)\ll\dot{M_a}$ is assumed constant averaged over durations of accretion and quiescence. Observationally, this rate driven by angular momentum loss by gravitational radiation from the binary is estimated through setting the measured time-averaged bolometric X-ray flux $\left<F_{X}\right>$ equal to the predicted value $(GM_{\rm NS}/(c^2 R_{\rm NS}))\dot{M}/(4\pi d^2)$~\cite{Galloway:2006ad}. The deep crustal heating power is given by~\cite{Yakovlev:2002ti} 
\bea
L_{\rm dh}&=&Q\times \frac{\dot{M}}{m_{\rm N}} \,\nn
&\approx& 6.03 \ee{33} \left(\frac{\dot{M}}{10^{-10}\,\mathrm{\Msolar~yr}^{-1}}\right) \frac{Q}{\rm MeV} \,\mathrm{erg~s}^{-1}~~
\label{eqn:dch}
\\
L_{\rm dh}^{\infty}&=&L_{\rm dh} \sqrt{1-2\,GM/(c^2 R)}\,
\label{eq:dch0}
\eea
where $m_{\rm N}$ is the nucleon mass, and $Q$ is the total amount of
heat released per one accreted nucleon (typically $Q\sim 1-2 \,\rm
{MeV}$, but values twice as large have been obtained for nucleon-nucleon interactions with a large nuclear symmetry energy~\cite{Steiner12dc}). Compared to \Eqn{eqn:Lum-ph}, \Eqn{eq:dch0} has only one redshift factor because the second is embedded in the definition of $\dot{M}$. For isolated neutron stars, $Q$ vanishes and the temperature evolution is dominantly cooling through energy loss of neutrino and photon emissions.

Starting from an arbitrary initial state, after a few thousands of cycles between accretion and quiescent state a thermal equilibrium is eventually reached independent of  the heat capacity of the star~\cite{Yakovlev:2002ti,Yakovlev:2003ed,Yakovlev:2004iq,Wijnands:2012tf}. Using the evolution code of temperature profiles in a spherically symmetric neutron star \footnote{http://www.astroscu.unam.mx/neutrones/NSCool} that solves the heat transport and hydrostatic equations including general relativistic effects~\cite{page1997thermal}, we perform numerical 
simulations and obtain the steady-state solution to~\Eqn{eqn:hc} which 
gives theoretical predictions on the $L_{\ga}^{\infty}-\dot{M}$ diagram 
(heating curves).

We do not consider prolonged and intense accretion phase (years to
decades rather than weeks to months) as in quasi-persistent low-mass
X-ray binaries (LMXBs), where stellar interiors are no longer maintained 
in thermal equilibrium. The steady luminosity decrease of these sources 
into quiescence can provide information on crust composition and properties~\cite{Roggero16,degenaar17,parikh16,Deibel:2016vbc} and limit the core heat capacity~\cite{Cumming:2016weq} (in some cases even for shorter outbursts the crust can come out of equilibrium, see~\cite{Degenaar:2013cbh, Waterhouse:2015kgo}).

\subsection{Heat-blanketing envelope and light elements}
\label{sec:eta}

The relation between the surface temperature $\ts$ and the internal
temperature $\ti$ is known by studying the uppermost envelope
($\rho\lesssim\rho_{b} \equiv10^{10}\, \rm{g\, cm^{-3}}$) as a thermal insulator separating the hot interior from the colder surface. Above the envelope lies the atmosphere ($\rho\lesssim 1\, \rm{g\, cm^{-3}}$) 
where $\ts$ is measured. The envelope is about one hundred meters 
thick with a strong temperature gradient, and the temperature at its 
bottom $\tb$ can be estimated in a simple formula~\cite{gudmundsson1983structure} 
\beq 
T_s \simeq10^{6} K \times \left(\frac{T_b}{10^{8} K}\right)^{0.5+\al}
\label{eqn:Ts-Tb}
\eeq
with $\al\ll 1$. At densities $\rho\gtrsim\rho_{b}$ the neutron star interior is nearly isothermal with a uniform internal temperature (including general relativistic effects) $\ti=T(r) \e^{\Phi(r)}=\tb$ owing to large thermal conductivities of degenerate matter.

The precise $\ts(\tb)$ relationship depends on the chemical composition of the envelope. Light elements like H or He considerably increase the thermal conductivity and result in a higher surface temperature than a heavy element. The thickness of the light-element layer is characterized as
\beq
\eta=g_{14}^2 \,\De M_{\rm le}/M
\label{eqn:eta}
\eeq where the surface gravity $g=GM/(c^2 R^2 \sqrt{1-r_g /R})$ is given in units of $10^{14}\, \rm{cm^2 \, s^{-1}}$ and $r_g$ is the Schwarzschild radius. The most massive light-element layer is limited around $\De M_{\rm le}\simeq 10^{-6}\,\Msolar$ ascribed to pycnonuclear reactions above $10^{10}\, \rm{g\, cm^{-3}}$. In Ref.~\cite{Brown:2002rf} the thickness of the helium layer is parametrized instead using the column density $y$. In this work, we ignore potential effects induced by mixtures of light elements in the envelope~\cite{Beznogov:2016ejn} and other modifications to the $\ts(\tb)$ relationship in the case of strong magnetic fields.

\subsection{Neutrino emission mechanisms}
\label{sec:nu}

\subsubsection{Main processes and equations of state}
\label{sec:nu_eos}
The total neutrino luminosity from the interior of the star~\Eqn{eqn:Lum-nu} relies on individual neutrino emissivities and their density and temperature dependence. Neutrino emission in nuclear matter mainly proceeds through bremsstrahlung of nucleon-nucleon, modified Urca, Cooper-pair breaking and formation (PBF), and direct Urca processes. The fastest cooling mechanism of all is direct Urca, of which the neutrino emissivity can be around six orders of magnitude larger than those of other mechanisms with vastly different temperature dependence (see Table~\ref{tab:nu-ems} for estimates and comparison), and it can only occur at densities with sufficiently high proton fraction~\cite{boguta81,Lattimer:1991ib}. In the present work we restrict ourselves to nucleonic degrees of freedom, and defer the investigation of neutrino emission rates for processes involving hyperons, quarks, pion and kaon condensates~\cite{Friman:1978zq,
Iwamoto:1980eb,Pethick:1991mk,prakash1994rapid,umeda1994neutron,
umeda1994nonstandard} etc. to future work.
\begin{table}[htb]
\begin{center}
\begin{tabular}{c c@{\quad} c}
\hline \\[-2ex]
Process & $\nu$ emissivity $(\mathrm {erg\,cm^{-3}\,s^{-1}})$ \\ [1ex]
\hline \\[-1.5ex]
Direct Urca & $\sim 10^{27}\, T_{9}^{6}$  \\[1ex]
Pair breaking & $\sim 10^{19} -10^{21}\, T_{9}^{7}$  \\[1ex]
Modified Urca & $\sim 10^{21}\, T_{9}^{8}$\\[1ex]
Bremsstrahlung & $\sim 10^{19}-10^{20}\, T_{9}^{8}$ \\[1ex]
\hline
\end{tabular}
\end{center}
\caption{Main processes of neutrino emission in nuclear matter.}
\label{tab:nu-ems}
\end{table} 
 
Table~\ref{tab:EoS} presents properties of a list of theoretical nuclear equations of state (EoSs) applied in our calculation; all these models fulfill the maximum mass constraint from two recently observed massive stars with $1.97\pm0.04\,\Msolar$ \cite{Demorest10} and $2.01\pm0.04\,\Msolar$ \cite{Antoniadis13}. Depending on the EoS and their masses, neutron stars with a central density higher than the direct Urca threshold undergo enhanced cooling. In contrast, the presence of enhanced cooling is prohibited artificially in the minimal cooling paradigm~\cite{Page04,Page:2009fu}, which is not employed in the current work.

\begin{figure}[htb]
\includegraphics[width=\hsize]{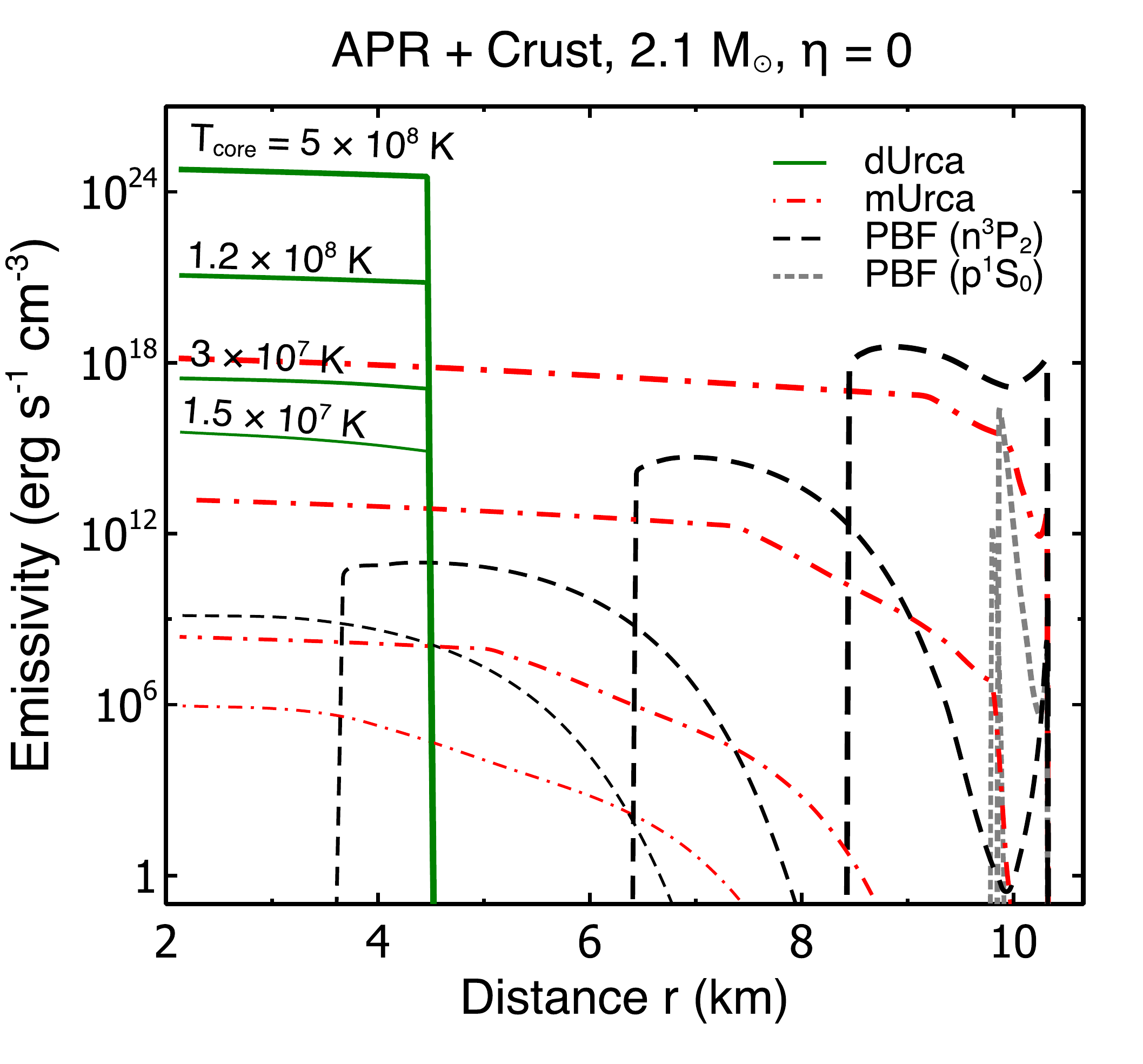}\\[2ex]
\caption{(Color online) Radial profiles ($r=0$ at the core) of the neutrino 
emissivity from various processes for a $2.1\,\Msolar$ APR neutron star 
at different core temperatures. Theoretical predictions from models 
``SFB-a-T73'' are employed for neutron singlet, neutron triplet and proton 
singlet pairing gaps (see text).
}
\label{fig:Q_r}
\end{figure}

In Fig.~\ref{fig:Q_r} displayed are emissivity profiles for four main neutrino cooling mechanisms in a $2.1\,\Msolar$ star based on the APR nuclear matter equations of state~\cite{akmal1998apr} with iron in the accreted  crust~\cite{Haensel:2007wy}; $\eta=0$ corresponds to the absence of a light-element layer in the envelope. Apparently one can see that above the onset mass $2.01\,\Msolar$, direct Urca neutrino emission in the core dominates cooling of the star.

In this calculation, pairing gaps of neutron/proton superfluidity are
described by $n\, \sfs$ ``SFB''~\cite{Schwenk:2002fq}, $n\, \sft$
``a''~\cite{Page04} and $p\, \sfs$ ``T73''~\cite{takatsuka73} models.
The core temperature $T_{\rm core}=\ti$ ranges from $5\ee{8}K$ to
$1.5\ee{7}K$, which plays a crucial role in determining the density
regions that are primarily affected by the PBF~\cite{Page:2010aw}
processes (see the next section). 

\begin{table}[htb]
\begin{center}
\begin{tabular}{c c c c@{\quad} c}
\hline \\[-2ex]
Property & APR & HHJ& SLy4& NL3\\[0.5ex]
\hline \\[-2ex]
Symmetry energy $S_0 $ (MeV)  & 32.6 & 32.0 & 32.0 & 37.3\\[0.5ex]
$L = 3 n_0 \, [ dS_0/dn ]_{n_0}$ (MeV) & 60 & 67.2 & 45.9 & 118.2\\[0.5ex]
Maximum mass of star ($\Msolar$)  & 2.18 & 2.17 & 2.05 & 2.77\\[0.5ex]
Direct Urca onset mass ($\Msolar$) & 2.01 & 1.87 & 2.03 & 0.82 \\[0.5ex]
Maximum density $n_{\rm max} \rm (fm^{-3})$ & 1.12 & 1.02 & 1.21 &
0.68 \\[0.5ex]
Direct Urca threshold $\nb^{\rm dU} \rm (fm^{-3})$ & 0.77 & 0.57 & 1.42 &
0.21 \\[0.5ex]
Radius of the heaviest star (km) & 10.18 &10.98& 9.96 & 13.65\\[0.5ex]
\hline
\end{tabular}
\end{center}
\caption{Calculated properties of the APR~\cite{akmal1998apr}, HHJ~\cite{heiselberg1999hhj} parametrized by~\cite{Kaminker:2014wna}, SLy4~\cite{Chabanat95} and NL3~\cite{lalazissis1997nl3} nuclear matter equations of state used in this work.
}
\label{tab:EoS}
\end{table} 
 
\begin{figure*}[htb]
\parbox{0.5\hsize}{
\includegraphics[width=\hsize]{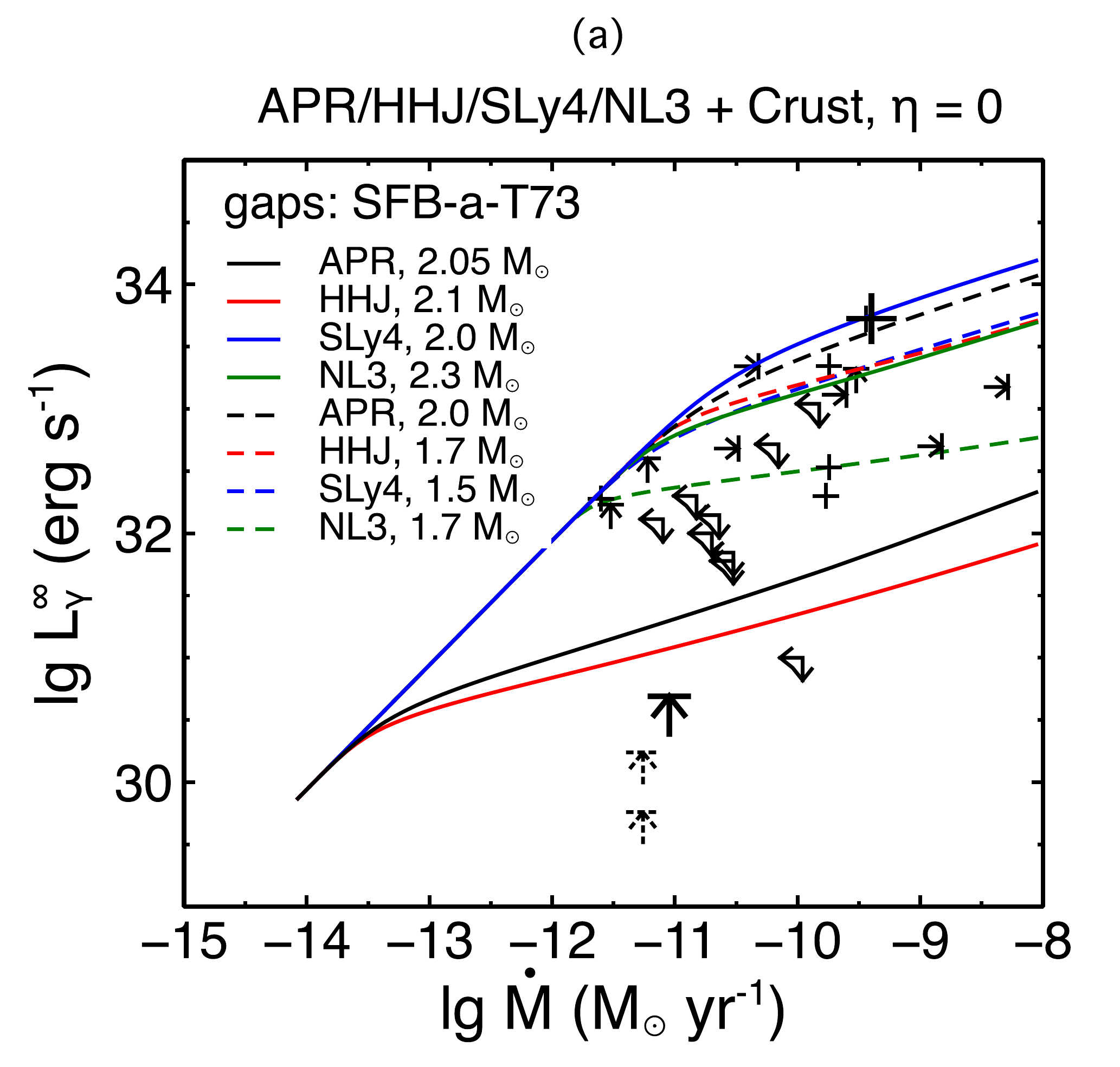}\\[-1ex]
}\parbox{0.5\hsize}{
\includegraphics[width=\hsize]{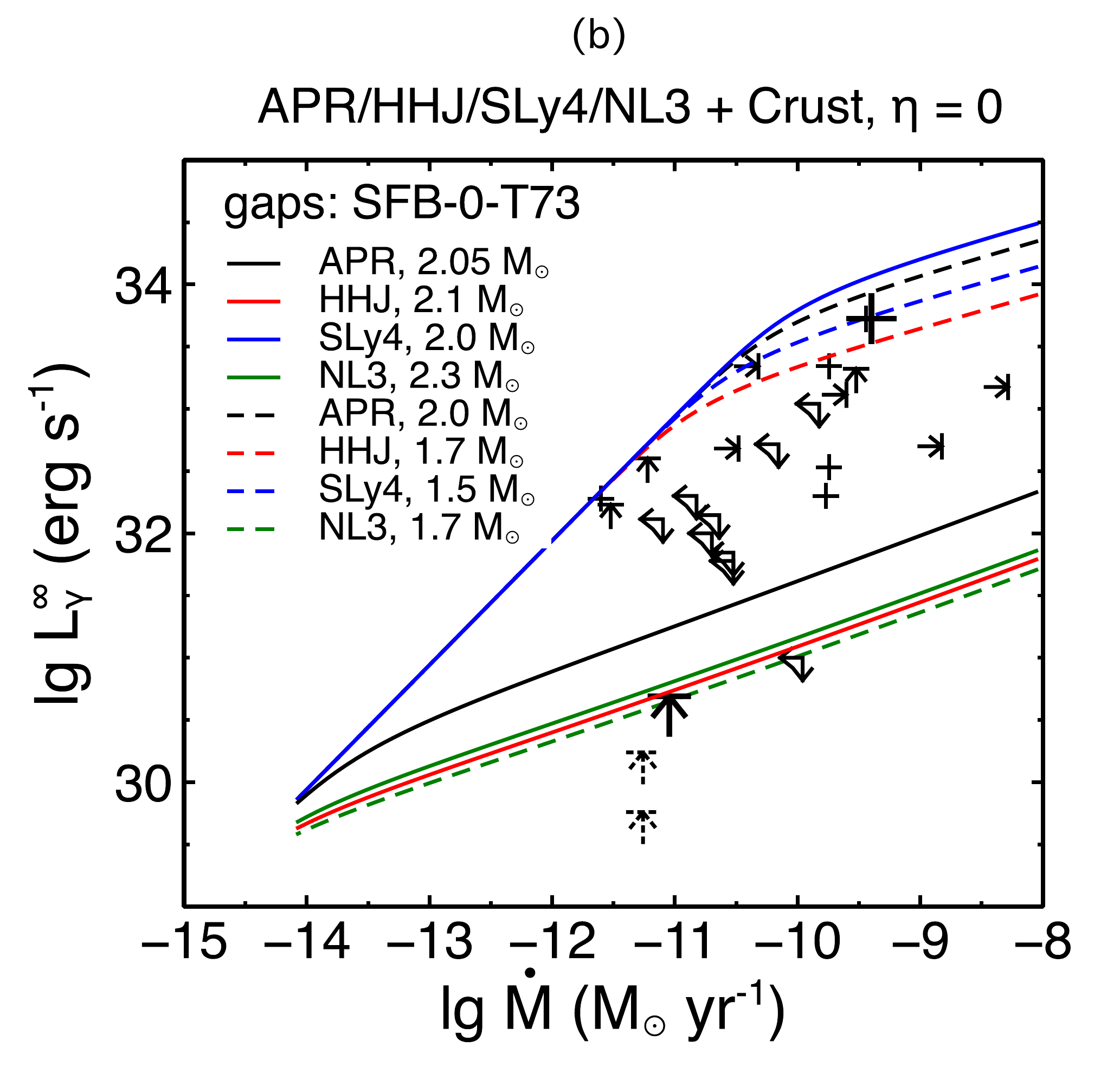}\\[-1ex]
}
\caption{(Color online) Panel (a): a series of selected heating curves $\mathrm{log}_{10} L_{\ga}^{\infty} - \mathrm{log}_{10}\dot{M}$
based on EoSs in Table~\ref{tab:EoS} with superfluidity gap models 
``SFB-a-T73'' and observational data taken from Refs.~\cite{Heinke10, 
Beznogov:2014yia}; two dashed arrows 
represent old data for the coldest star in SAX J1808~\cite{campana2002xmm} 
to be replaced by the most recent one (enlarged solid arrow) with distance 
estimate updated~\cite{Galloway:2006ad}, and the line cross with highest 
luminosity stands for the hottest star in Aql X-1. Panel (b): similar figure 
for ``SFB-0-T73'' gap models.
}
\label{fig:hc-vary-eos-Tc}
\end{figure*}

\subsubsection{Superfluidity}
\label{sec:sf}

Pairing gaps of stellar superfluid have primary effects on the thermal 
evolution of neutron stars. First, there are strong suppressions (by a 
factor $\sim e^{-2\De/T}$ where $\De$ is the pairing gap) of neutrino 
emissivities and the specific heat at temperatures far below the critical
temperature ($T\ll\tc$). Second, at {ambient temperatures slightly below 
the critical temperature ($T\lesssim\tc$), the specific heat is modified 
and the PBF neutrino emission process is triggered~\cite{Flowers:1976ux,
levenfish1994specific,levenfish1994suppression,levenfish1996standard}. 
The relationship between the zero temperature gap and $\tc$ is 
approximated by the BCS theory as $\De(T=0)\simeq1.75\,\tc$~\cite{BCS57}. As the specific heat effect is negligible in an accreting neutron star under thermal equilibrium $\mathrm{d}\ti/\mathrm{d}t=0$ (see equivalently~\Eqn{eqn:hc}), we concentrate on the noticeable change in neutrino emissivity induced by superfluidity.

In principle, dense-matter equations of state and pairing gaps are described by the same Lagrangian within a unified framework; however,
self-consistent theoretical calculations are a formidable task since the pairing gap is exponentially sensitive to the coupling potential. Therefore, it is common to search for reasonable combinations of EoSs and gap models so that one can test restrictions jointly from observational data.

To give a reliable estimate of how superfluidity can alter the thermal 
states, we adopt ``SFB-a-T73'' gap models as mentioned above; note that the $n\sft$ gaps are anisotropic and angle-averaged values are given~\cite{Page04}, extending to inner core regions. Calculations of proton $\sfs$ superfluid critical temperatures $T_{\rm cps}$ from the ``T73'' model show that the gap vanishes at densities above $k_{F \rm{p}}\simeq 0.8\, \rm{fm}^{-1} (\rho_b\simeq 4.5\times10^{14}\, \rm{g\, cm^{-3}}) $, which is still below the onset density for nucleonic direct Urca process in the APR equation of state (Table~\ref{tab:EoS}). Stated differently, in this example as can be seen from Fig.~\ref{fig:Q_r}, PBF neutrino emission in the $p\sfs$ channel is detached from direct Urca (most efficient cooling) and the suppression of the latter is regulated only through $n\sft$ pairing. Models of proton singlet gaps that extend to densities near the direct Urca threshold have been discussed in other works, see e.g. Ref.~\cite{Yakovlev:2004iq} or more recently Ref.~\cite{Taranto:2015ubs}.

Alternatively, to describe the $n\sft$ gap in the inner core and the $p\sfs$ gap in the outer core, we use a phenomenological Gaussian approximation~\cite{Beloin17} which has six parameters $[T_{\rm cnt}^{\rm max},k_{F\rm n}^{\rm peak},\De k_{F\rm n}; T_{\rm cps}^{\rm max},k_{F\rm p}^{\rm peak},\De k_{F\rm p}]$. This parametrization will be used in the statistical analysis in Sec.~\ref{sec:stat}. Neutron $\sfs$ pairing is present in the neutron star crust, with critical temperatures too large to be constrained by the observational data. For consistency, we keep the SFB model throughout our calculations. Unlike with the combination of the T73 proton gap and the APR EoS described above, some of the parametrizations with large values of $k_{F\rm p}^{\rm peak}$ or $\De k_{F\rm p}$ will have proton gaps which operate inside the region allowed by the direct Urca process.

\begin{figure*}[htb]
\parbox{0.5\hsize}{
\includegraphics[width=\hsize]{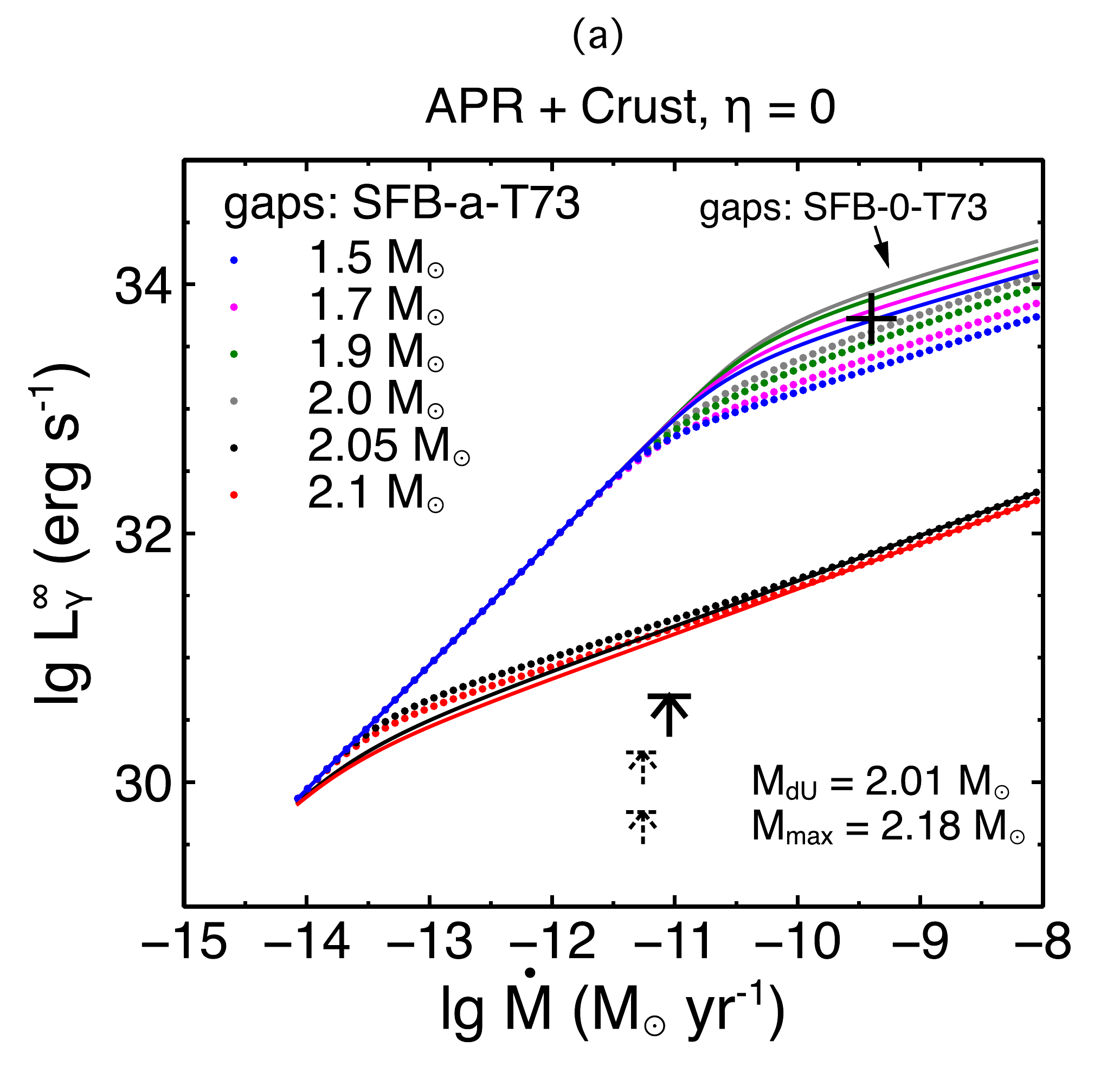}\\[-2ex]
}\parbox{0.5\hsize}{
\includegraphics[width=\hsize]{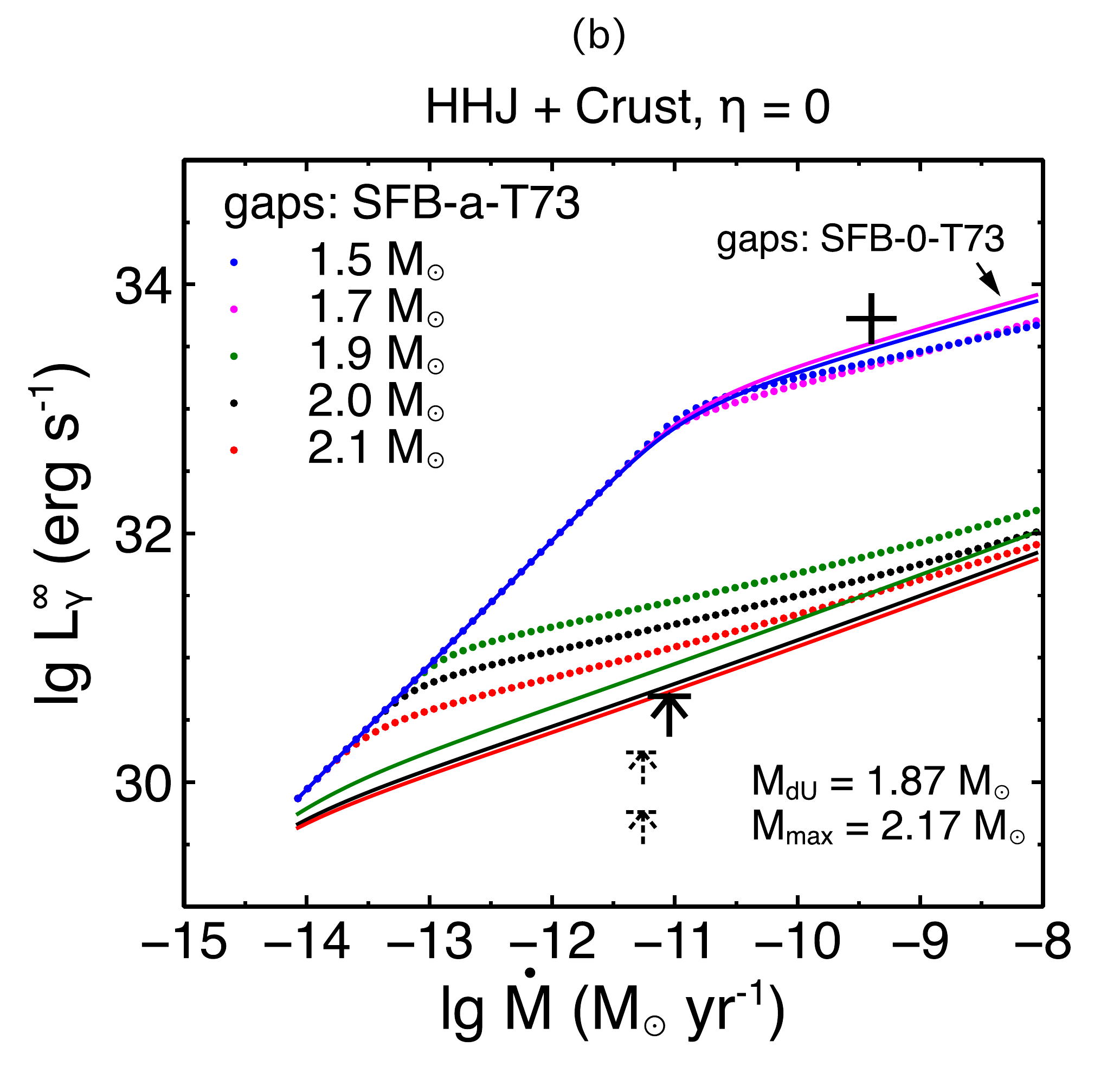}\\[-2ex]
}

\caption{(Color online) Exemplary heating curves $\mathrm{log}_{10} L_{\ga}^{\infty} - \mathrm{log}_{10}\dot{M}$ for APR/HHJ EoS in 
comparison with luminosity and accretion rate measurements of the 
hottest (line cross, Aql X-1) and coldest (arrow, SAX J1808) sources; 
light elements (H or He) are not taken into account. See Fig.~\ref{fig:hc-vary-eos-Tc} panel (a) for more EoSs and observational data.
}
\label{fig:heating-curves-1}
\end{figure*}

\begin{figure*}[htb]
\parbox{0.5\hsize}{
\includegraphics[width=\hsize]{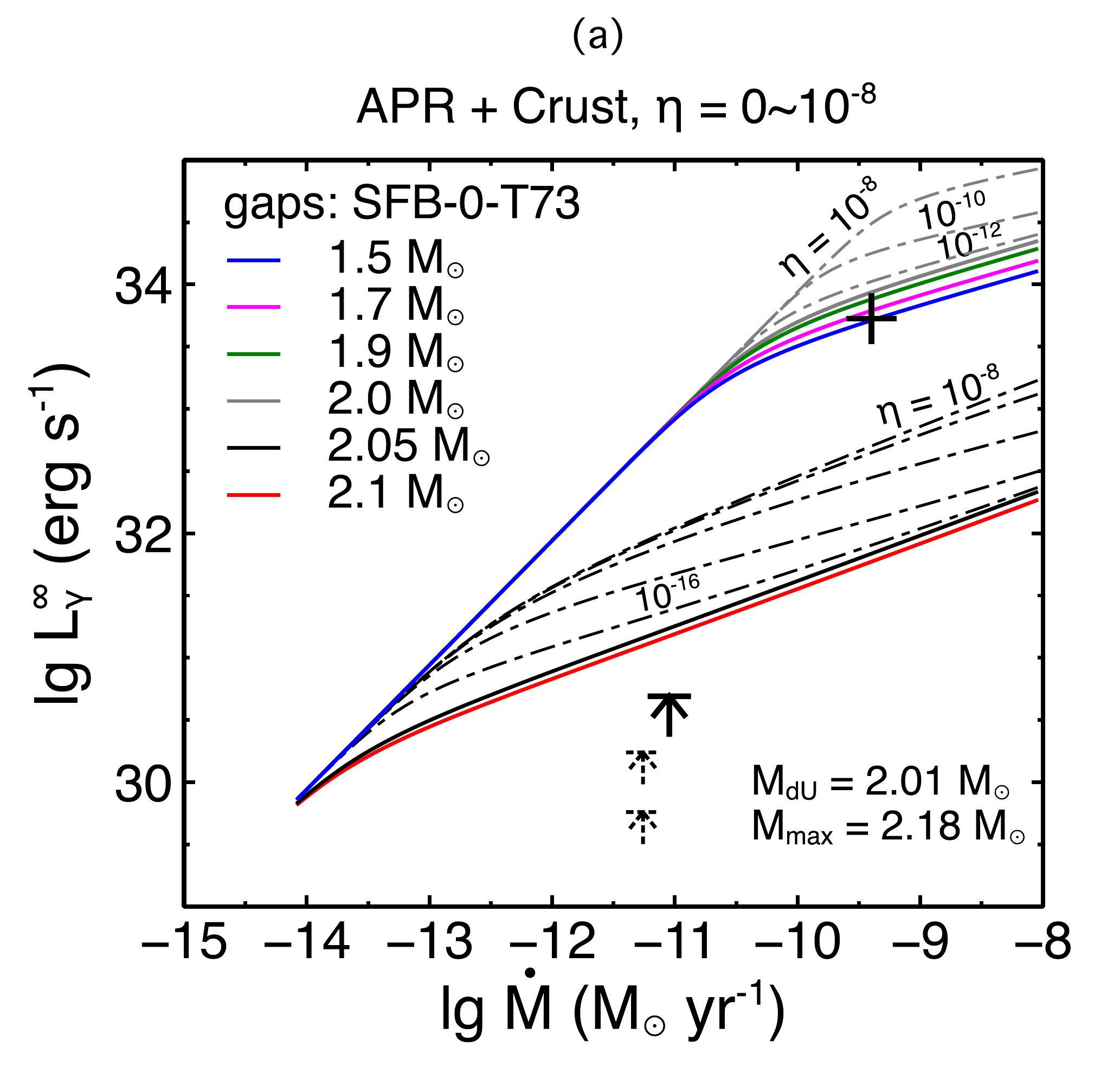}\\[-2ex]
}\parbox{0.5\hsize}{
\includegraphics[width=\hsize]{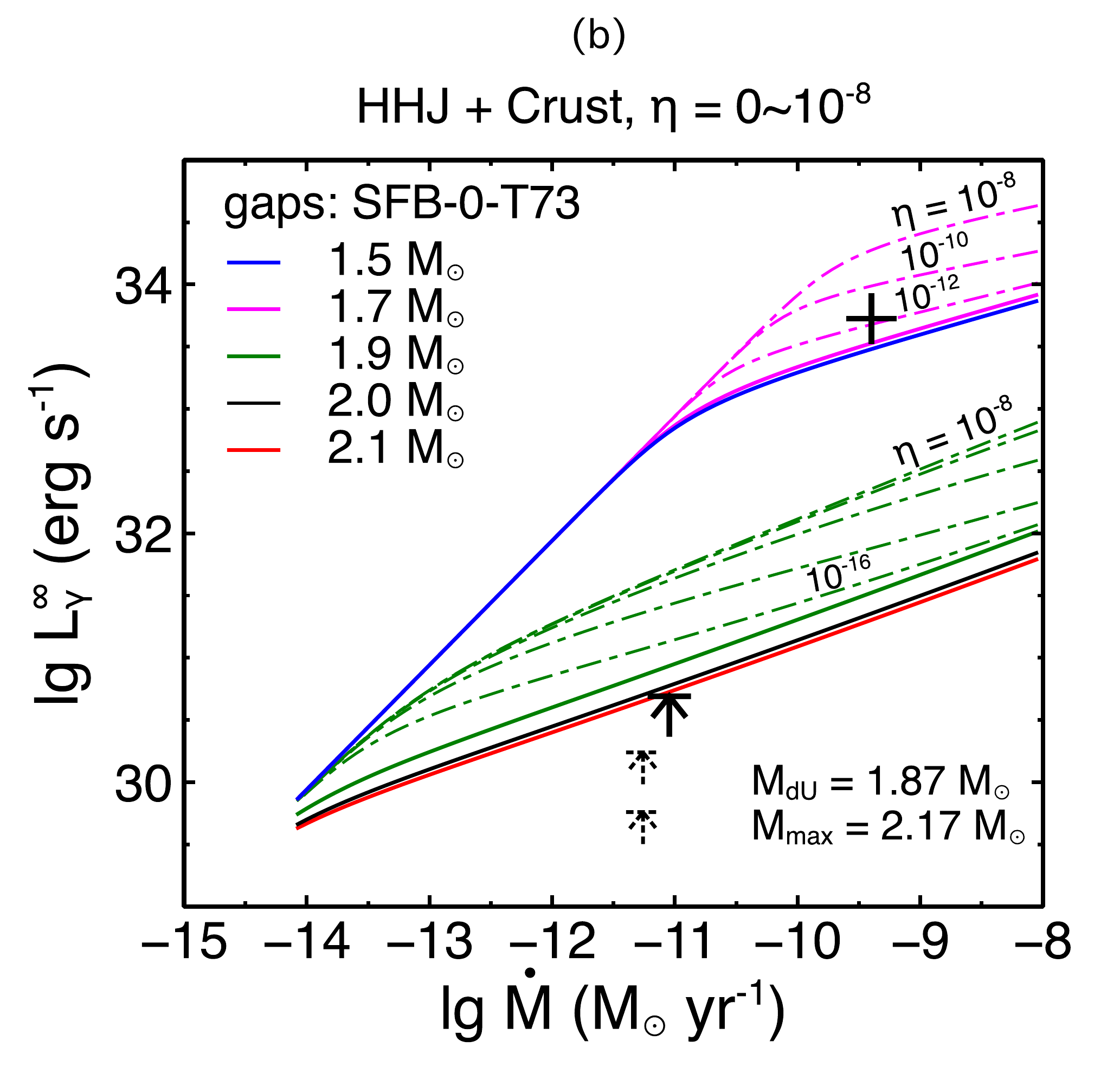}\\[-2ex]
}

\caption{(Color online) Heating curves $\mathrm{log}_{10} L_{\ga}^{\infty} - \mathrm{log}_{10}\dot{M}$ for APR/HHJ EoS, with superfluidity 
models ``SFB-0-T73'' (vanishing $n \sft$ gap). Effects from light-element 
layer in the envelope (see \Eqn{eqn:eta}) are shown for two different masses 
(APR: $2.0\,\Msolar, 2.05\,\Msolar$; HHJ: $1.7\,\Msolar, 1.9\,\Msolar$) below 
and above the direct Urca onset.
}
\label{fig:heating-curves-eta}
\end{figure*}

\section{Observations}
\label{sec:obs}
Displayed in Fig.~\ref{fig:hc-vary-eos-Tc} are measurements of, or limits 
on the quiescent surface photon luminosity $L_{\ga}^{\infty}$ and the 
averaged mass transfer rate $\dot{M}$ of X-ray transients harboring 
neutron stars adapted from earlier work~\cite{Heinke:2008vj,Heinke10, Beznogov:2014yia}. It is worth mentioning that for simplicity the uncertainty of distance measurements by a factor of 1.5 as in~\cite{Heinke:2006ie,
Heinke:2008vj,Heinke10} has not yet been included. In principle, each data point should reside on a heating curve generated for a particular composition. 

To date the coldest neutron star located in SAX J1808.4-3658~\cite{Heinke:2008vj,Galloway:2006ad,campana2002xmm} very likely carrying large moment of inertia~\cite{Patruno:2016vqc} and the hottest one in Aql X-1~\cite{Rutledge:2001kc,Campana:2003cz,tomsick2004low,Heinke:2006ie} are two most distinctive candidates of which future observations are promising to impose even more stringent limitations on theory.

Following this strategy, we perform consistent calculations of thermal
evolution in transients. Confronted with the observational data of 24
systems are the aggregated results of our numerical calculation for a
collection of nuclear matter EoSs and neutron star masses, with
superfluidity gap models ``SFB-0-T73'' (``0'' refers to zero $n \sft$ gap 
in contrast to model ``a'' for a mild triplet gap; see detailed discussion 
in Sec.~\ref{sec:hc}).

\section{Results and discussion}
\label{sec:result}

\subsection{Heating curves (APR/HHJ EoS)}
\label{sec:hc}

The solutions to the thermal balance equation \eqn{eqn:hc} give heating 
curves $L_{\ga}^{\infty}(\dot{M})$ depicting how the surface photon 
luminosity during quiescence depends on the mean accretion rate.

We adopt two EoSs, (i) APR EoS and (ii) HHJ EoS, presented in Table
\ref{tab:EoS}, and calculate heating curves progressively varying the
neutron star mass as shown in Fig.~\ref{fig:heating-curves-1}. Iron
envelope is assumed, with the total amount of heat released per one
accreted nucleon $Q= 1.45\,\rm {MeV}$ in \Eqn{eqn:dch}, and there is
no accreted light-element layer ($\eta=0$). For the superfluidity gap
models, we select two sets of combination, ``SFB-a-T73'' and ``SFB-0-T73'', in the attempt to figure out the effect of $n\sft$ neutron triplet gap. Model ``a'' stands for a mild triplet gap with $\tc\simeq 10^{9}K$ %\old{while vanishing neutron triplet gap ``0'' has zero effect in terms of suppressing neutrino emissivity.} 
and refers to the dotted curves. The thin solid curves represent the results with vanishing neutron triplet gap, which has zero effect in terms of suppressing neutrino emissivity.

Analogous to cooling isolated neutron stars, accreting neutron stars
may be operating in the photon-emission regime or the neutrino-emission 
regime. At sufficiently low accretion rate, all heating curves merge into a 
single, universal curve which denotes the limiting case $L_{\nu}\ll L_{\ga}\approx L_{\rm dh}$. 
In this photon-emission regime, the surface luminosity is solely determined 
by the accretion rate, irrespective of the neutron star interior composition and structure. The latter era where $L_{\ga}\ll L_{\nu}\approx L_{\rm dh}$, is realized at higher accretion rate and very sensitive to the internal structure (and essentially the neutrino emission mechanism). Observational data suggest that the majority of accreting neutron stars in transients are in the neutrino-emission regime, hence their thermal states offer an exceptional tool to probe the underlying physics of dense matter.

We see clearly that heating curves are separated by the direct Urca onset 
mass, above which fast neutrino emission is switched on, therefore the core temperature (and the surface temperature/luminosity) drops dramatically. As a consequence, massive stars with enhanced cooling mechanism can naturally explain the fact that even at quite high accretion rate some very cold sources have been detected, and observation of these cold sources in turn probe the densest matter realized in neutron star interiors. Low-mass and intermediate-mass stars instead, with the less efficient modified Urca process as the dominant cooling mechanism, stay warmer. Furthermore, as discussed in Sec.~\ref{sec:eta}, finite amount of light elements in the envelope increases the thermal conductivity, which implies a higher surface luminosity at given core temperature and help 
push the heating curves upwards. This is best demonstrated in Fig.~\ref{fig:heating-curves-eta}, where $\eta$ that characterizes the amount of light elements is varied from 0 to $10^{-8}$ to generate different heating curves for given masses. It is obvious that the inclusion of light elements has basically two effects: first, it relieves the tight constraint from the hottest star, and if transients with even higher luminosity were to be observed we would need resort to other explanations e.g. deep crustal 
heating might be more powerful than we have assumed~\cite{Steiner12dc}; second, comparing to Fig.~\ref{fig:hc-vary-eos-Tc} where several X-ray transients fall into a ``gap'' between heating curves within a small mass range, one can see that some of them can now be successfully explained by adjusting the light-element amount, while the gap shrinks but still exists.

Our present results are qualitatively consistent with the literature~\cite{
Heinke:2008vj,Yakovlev:2002ti,Yakovlev:2003ed}. The standard cooling scenario agrees with predictions of low-mass and intermediate-mass stars, taking into account the flexibility of varying light-element amounts. However, the coldest star in SAX J1808.4-3658 cannot be explained without inclusion of additional fast cooling due to direct Urca process involving nucleons and/or hyperons. It is noteworthy that by virtue of general relativistic effects, the luminosity does not always decrease 
monotonically with the neutron star mass (it in fact grows with mass before the direct Urca is switched on); therefore the hottest star has no definite link to the lightest mass (see also in Ref.~\cite{Ofengeim:2016rkq}).

Additionally, comparison among heating curves calculated from different superfluidity parameters infers the preference of small $n\sft$ gaps. The mild triplet gap ``a'' has trouble explaining both hot and cold stars, although one can tune the light-element amount to achieve the high luminosity e.g. of the neutron star in Aql X-1. On the one hand, the hot stars with core temperatures relevant to the critical temperature for neutron triplet gap undergo PBF neutrino emissions, which lowers the heating curves compared to the case of no neutron superfluid. On the other hand, at the 
core of cold massive stars, the suppression of direct Urca prevents them from cooling to temperatures as low as that of SAX J1808.4-3658. EoS-dependence is also important, as is shown in Fig.~\ref{fig:hc-vary-eos-Tc} in that it designates the minimum mass associated for direct Urca process. Take the NL3 EoS as an example, where direct Urca kicks in at comparatively low density ($M_{\rm dU}=0.82\,\Msolar$). It is not surprising that once neutrons form superfluid the suppression of enhanced cooling remain drastic for most neutron star masses, which gives rise to the substantial variation between NL3 heating curves in the left and right panel 
of Fig.~\ref{fig:hc-vary-eos-Tc}. The HHJ EoS with zero neutron triplet gap ``0'' marginally suffices to explain the observational data of SAX J1808.4-3658.

\begin{figure}[htb]
\includegraphics[width=\hsize]{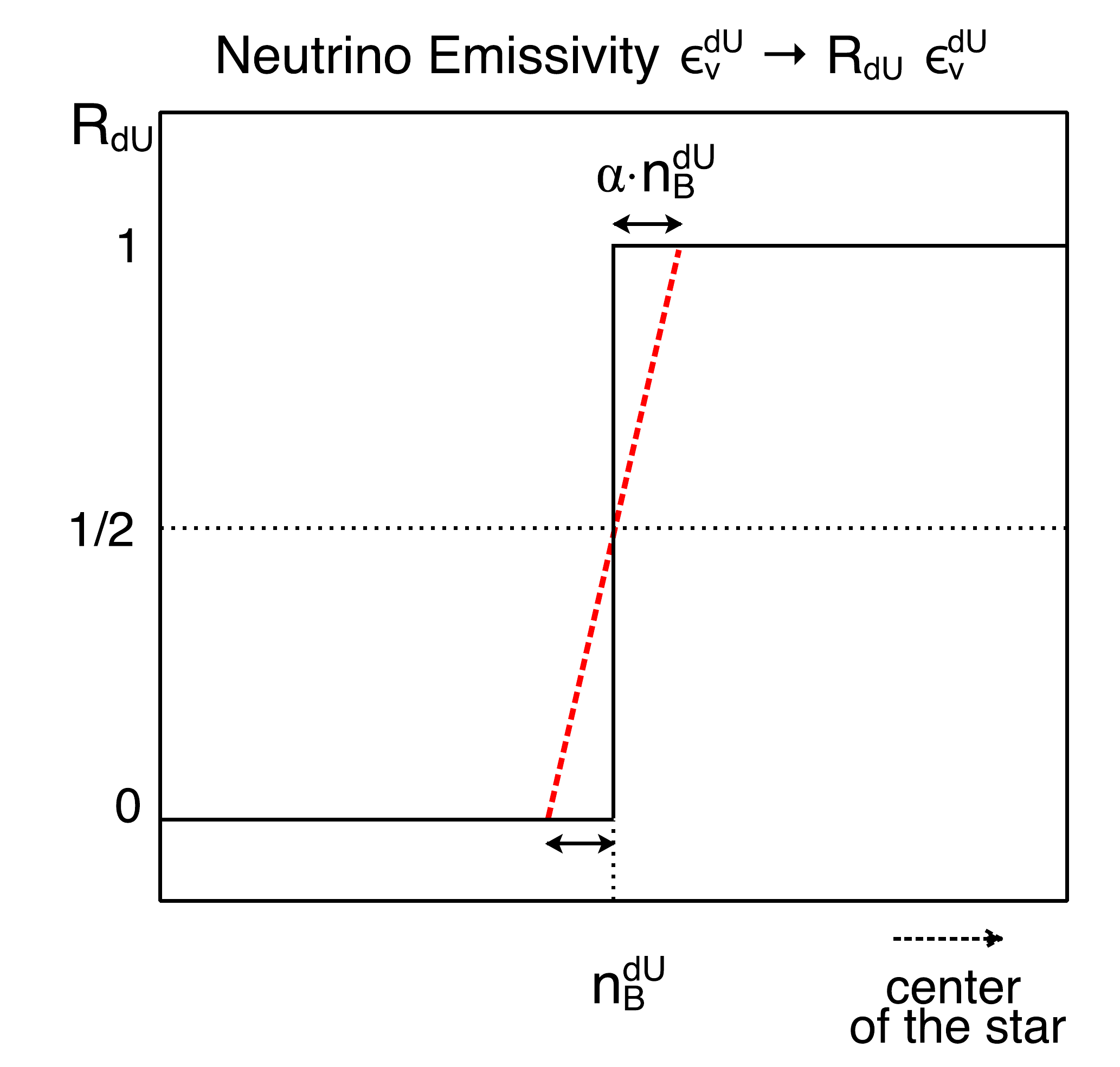}\\[-2ex]
\caption{Schematic plot of regulating the direct Urca emissivity near its threshold. See \Eqn{eqn:dU_mod} and text.
}
\label{fig:R_dU}
\end{figure}

\begin{figure*}[htb]
\parbox{0.5\hsize}{
\includegraphics[width=\hsize]{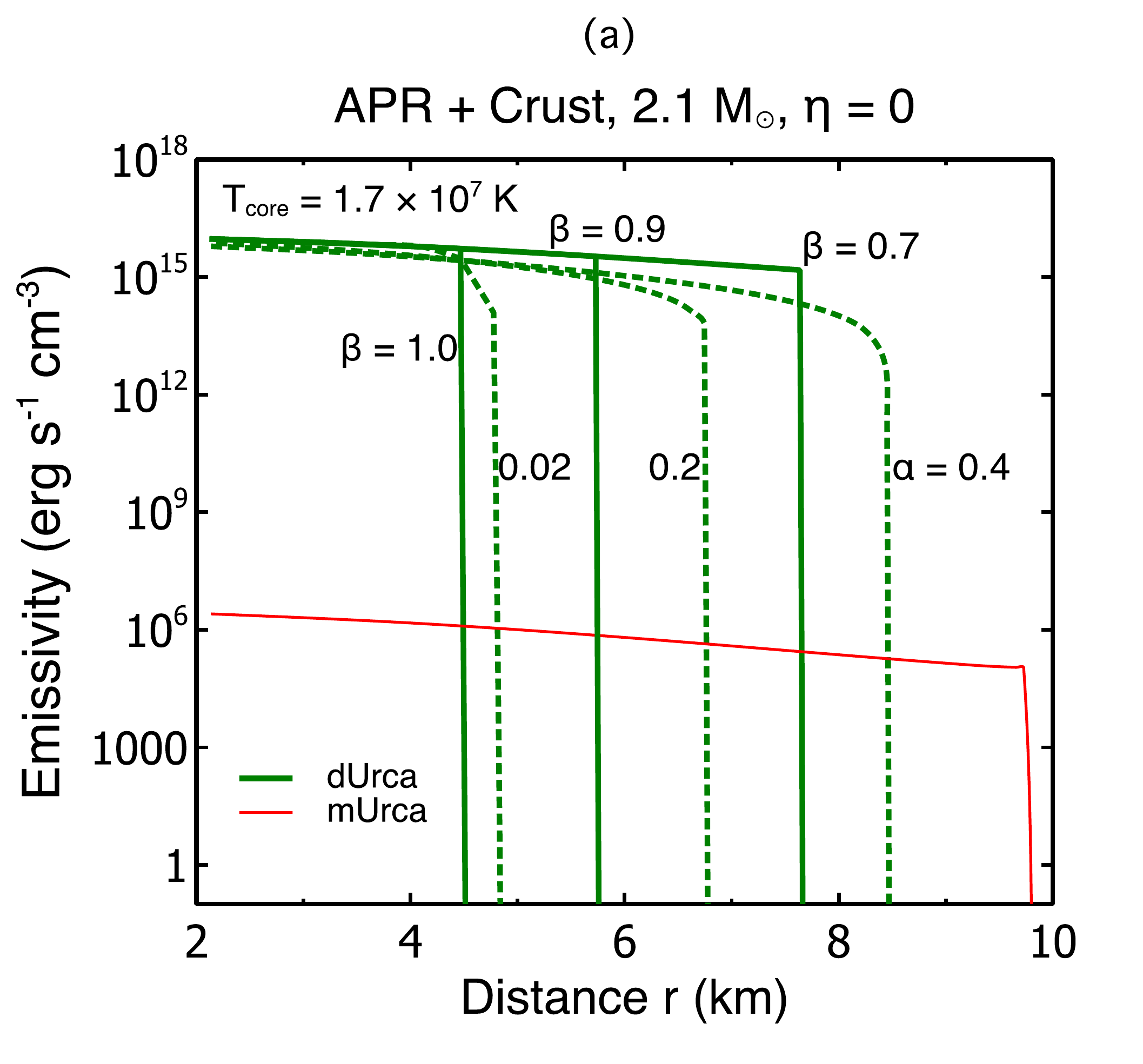}
}\parbox{0.5\hsize}{
\includegraphics[width=\hsize]{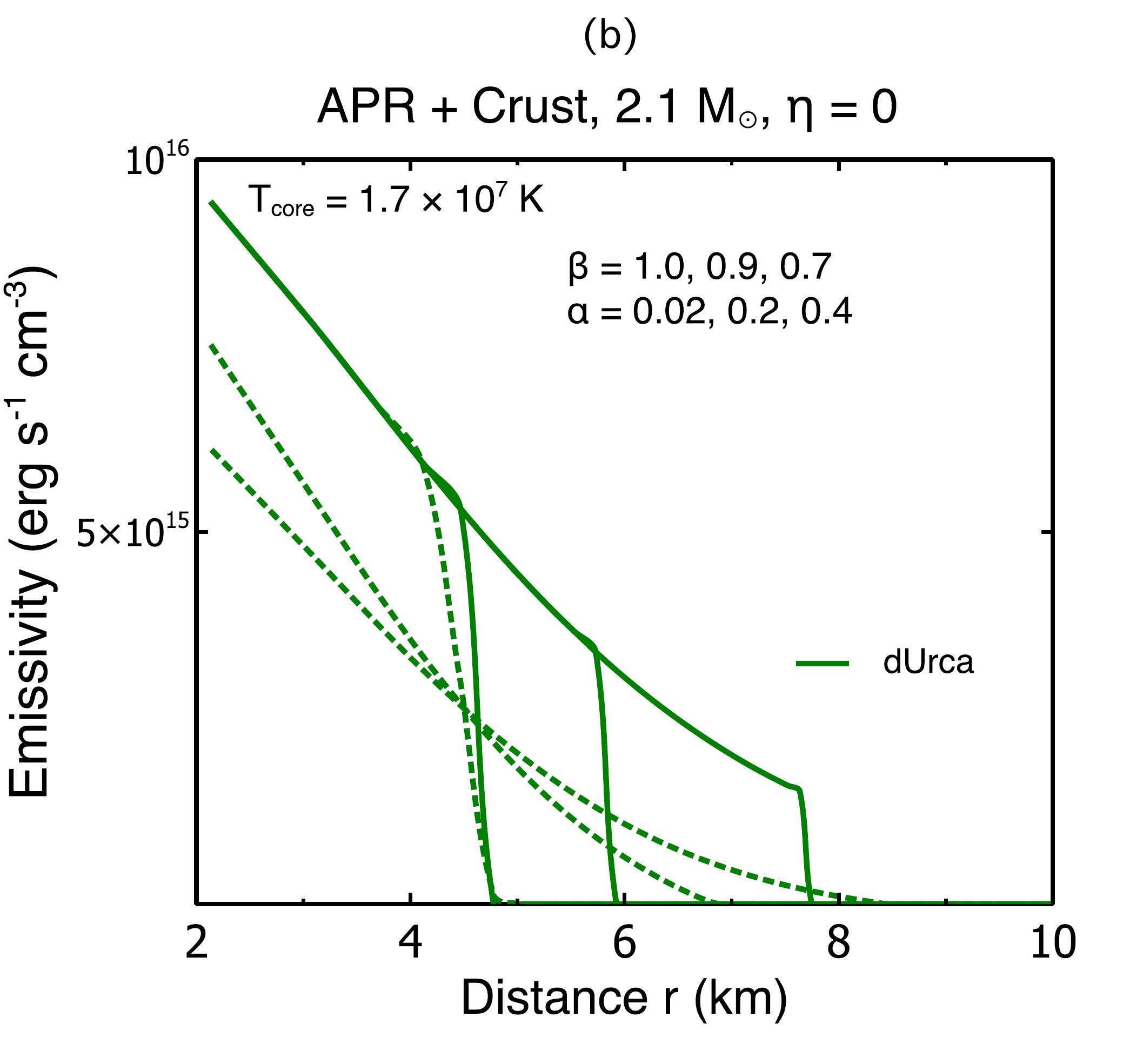}
}

\parbox{0.5\hsize}{
\includegraphics[width=\hsize]{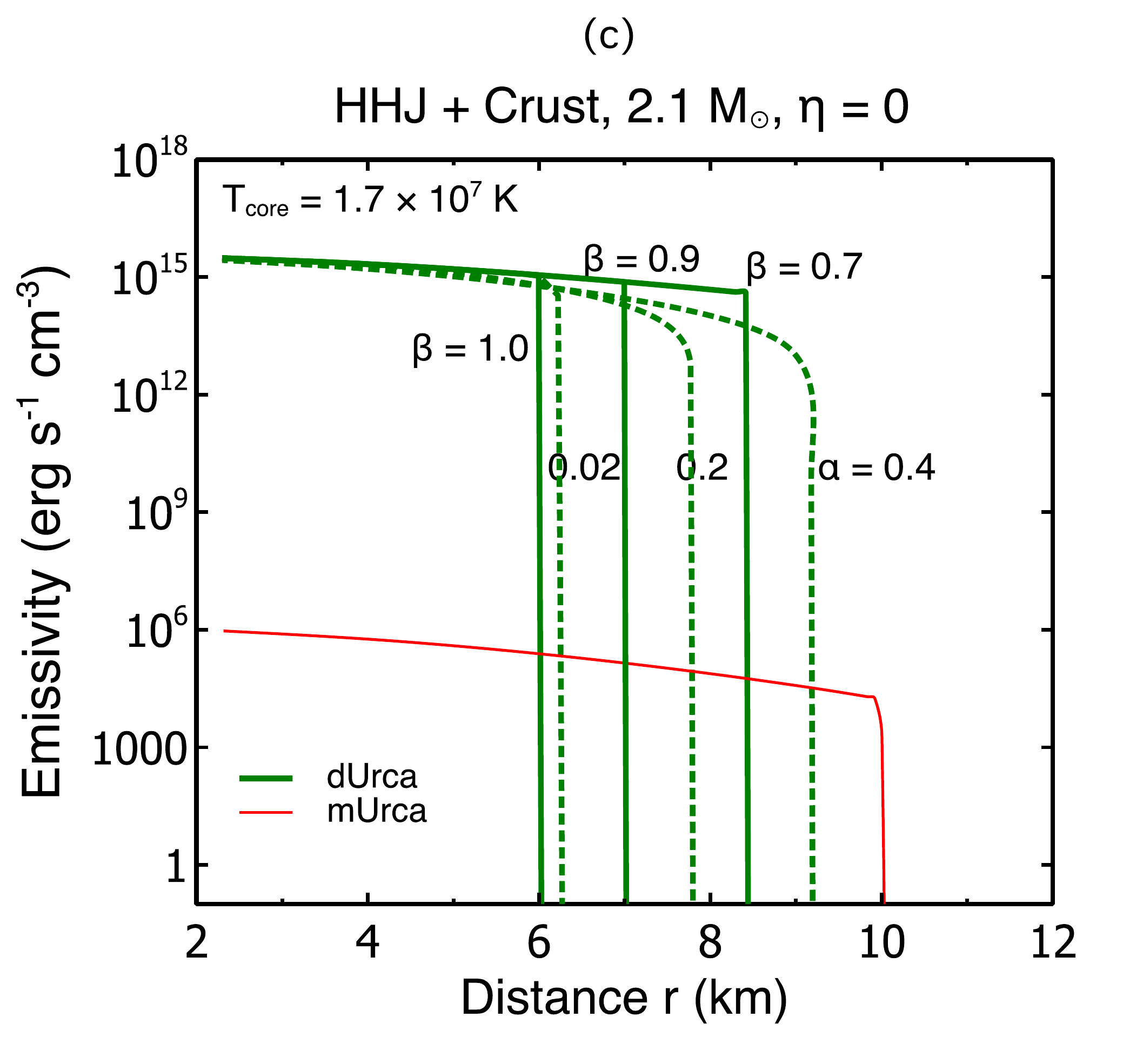}\\[-2ex]
}\parbox{0.5\hsize}{
\includegraphics[width=\hsize]{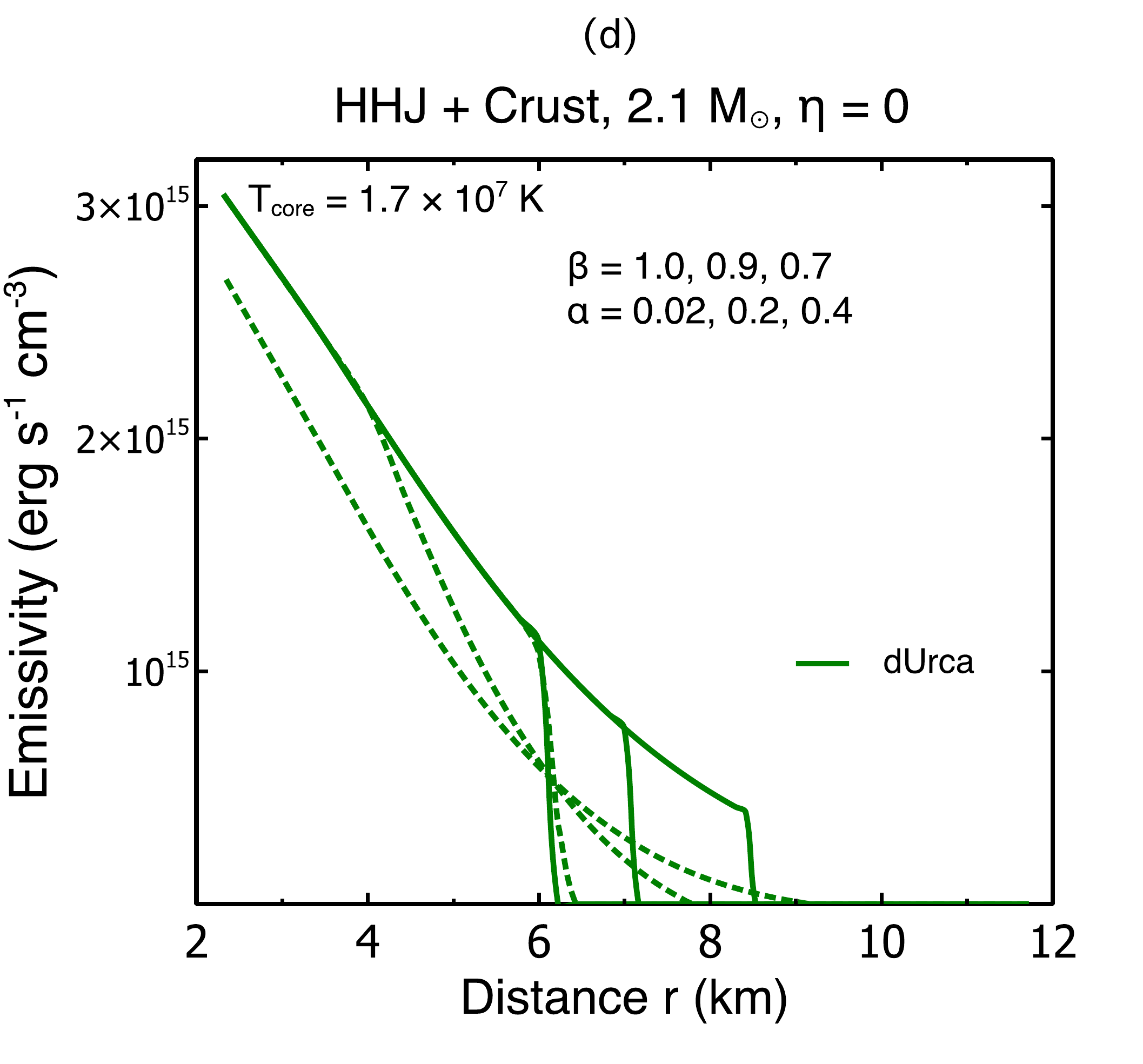}\\[-2ex]
}

\caption{(Color online) Neutrino emissivities plotted for a $2.1\,\Msolar$ 
star taking into account shifting and broadening of the direct Urca threshold (see Sec.~\ref{sec:dU-Sf}). LHS: logarithm scale; RHS: linear scale (direct 
Urca only).
}
\label{fig:emissivity-profiles}
\end{figure*}

\begin{figure*}[htb]
\parbox{0.5\hsize}{
\includegraphics[width=\hsize]{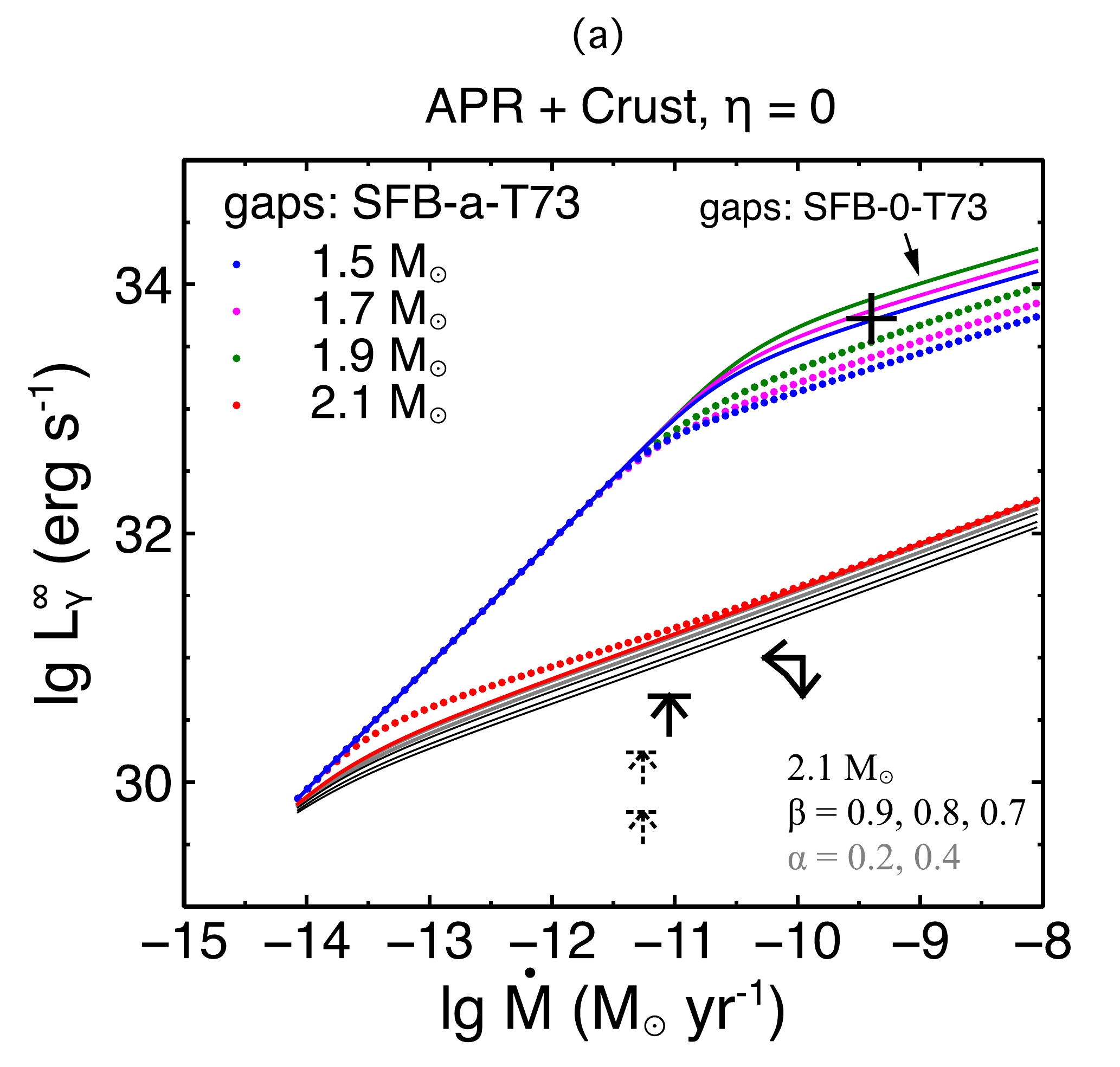}\\[-2ex]
}\parbox{0.5\hsize}{
\includegraphics[width=\hsize]{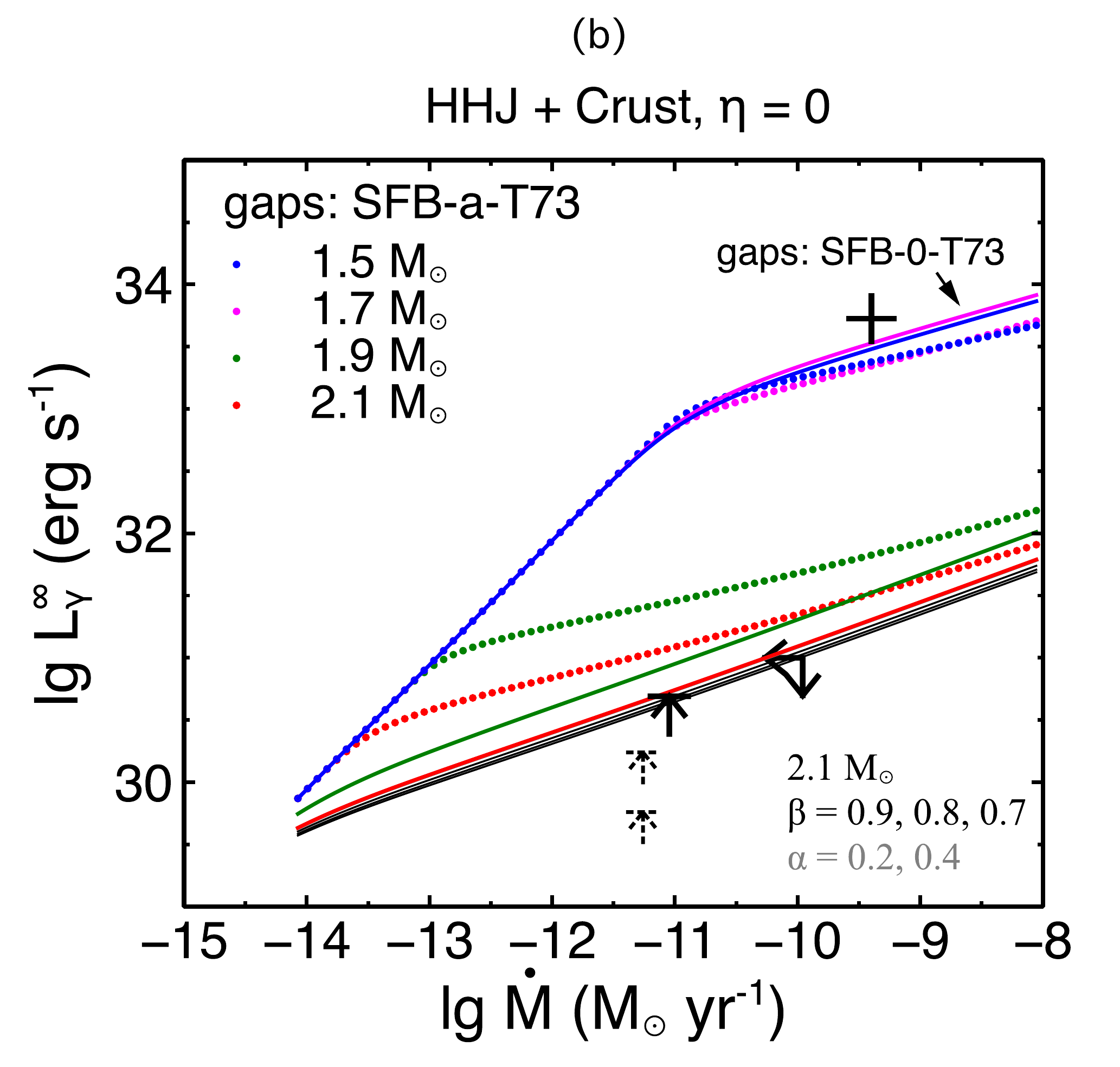}\\[-2ex]
}
\caption{(Color online) Introduce shifting and broadening of direct Urca process and 
recalculate the heating curves $\mathrm{log}_{10} L_{\ga}^{\infty} - \mathrm{log}_{10}\dot{M}$ as in Fig.~\ref{fig:heating-curves-1}. Double arrows represent the  measurements of 1H 1905+000 which are 
also compelling. See discussions in the text.
}
\label{fig:heating-curves-2}
\end{figure*}

\subsection{Direct Urca threshold}
\label{sec:dU-Sf}
The tight constraint from cold stars can be relieved by a lowering/broadening of the threshold for the enhanced cooling. We either replace the step function at the direct Urca onset $\nb^{\rm dU}$ by a linear interpolation of emissivities in a nearby region (see \Eqn{eqn:dU_mod} below), or introduce an arbitrary pre-factor in front of the threshold density. It has been pointed out that if the nuclear symmetry energy turns out to have a strong density dependence, direct Urca process will be triggered at lower densities than expected~\cite{Lattimer:1991ib,prakash1992rapid,
Steiner06hs}. Besides, the broadening of direct Urca threshold is a possible scenario in the context of superfluidity, pion polarization, thermal excitation or strong magnetic field~\cite{Yakovlev:2000jp}. Recent work~\cite{Beznogov:2014yia,Beznogov:2015ewa} introduced a similar phenomenological effect by smoothing the step function with a nonlinear and more elaborate control function.

Fig.~\ref{fig:R_dU} is an illustrative diagram showing the control parameter 
that modifies neutrino emissivities of the direct Urca process $\eps_{\nu}^{\rm dU}\to R_{\rm dU}\,\eps_{\nu}^{\rm dU}$ as a function of the baryon number density, where
\beq
R_{\rm dU} (s) = \left\{\!
\begin{array}{ll}
0 & s<(1-\al)\\[2ex]
\dsp\frac{1}{2}+\frac{s-1}{2\,\al}& (1-\al)\leq s<(1+\al)\\[2ex]
1 & s\geq(1+\al)
\end{array}
\right.\ 
\label{eqn:dU_mod}
\eeq with the scaling factor $s\equiv \nb/\nb^{\rm dU}$. The other option of lowering the direct Urca threshold $\nb^{\rm dU}\to \be\nb^{\rm dU}$ is also applied, as can be seen in Fig.~\ref{fig:emissivity-profiles}, where we demonstrate how radial profiles of neutrino emissivities from direct Urca and modified Urca processes, for a $2.1\,\Msolar$ APR star with core temperature $\ti=1.7\times 10^{7} K$, are altered by varying the broadening and shifting parameters $\al=0.02,0.2,0.4; \be=1.0,0.9,0.7$.

Implementing these phenomenological modifications to the direct Urca threshold, we can recalculate the $L_{\ga}^{\infty}-\dot{M}$ relation in Fig.~\ref{fig:heating-curves-1}. In Fig.~\ref{fig:heating-curves-2}, %\old{lower} 
grey and black thin lines represent updated heating curves with broadening/shifting effects for a $2.1\,\Msolar$ star; one more cold source in 1H 1905+000~\cite{
Jonker:2006td,jonker2007cold,Heinke:2008vj} is added for comparison. As one can read from the result, quiescent surface luminosities are further reduced for massive stars as expected: for APR EoS heating curves of the $2.1\,\Msolar$ star exhibit a maximal decrease in $L_{\ga}^{\infty}$ by $\sim39\%$ ($\beta=0.7$) and for HHJ by $\sim21\%$ ($\beta=0.7$) compared to their original values. In the case of HHJ EoS without light elements, cold sources such as SAX J1808.4-3658 and 1H 1905+000
can be well explained provided that neutron $\sft$ gap is vanishingly
small and direct Urca threshold is lowered/broadened to some extent.

\begin{table}[htb]
\begin{center}
\begin{tabular}{c c@{\quad} c}
\hline \\[-2ex]
Parameter & Limits \\[0.5ex]
\hline \\[-2ex]
$\nb^{\rm dU}, \al$          & $\nb^{\rm dU}(1-\alpha) \geq
0.16~\rm{fm}^{-3}$  \\[0.5ex]
$\eta_{\mathrm{Aql}}$          & $10^{-17}\sim10^{-7}$  \\[0.5ex]
$Q$          & $1 - 1.5 \, \rm{MeV}$  \\[0.5ex]
$T_{\rm cnt}^{\rm max}, T_{\rm cps}^{\rm max}$ & $10^{7}\sim10^{10}\,K$ \\[0.5ex]
$k_{F\rm n}^{\rm peak}, k_{F\rm p}^{\rm peak}$ & inside core of the
star with $M=M_{\mathrm{max}}$ \\[0.5ex]
$\De k_{F\rm n}, \De k_{F\rm p}$ &
ensure $\De k_{F} \leq k_{F}^{\mathrm{peak}}$ \\[0.5ex]
$M_{1808}, M_{\rm Aql}$    & ensure $M \leq M_{\mathrm{max}}$
\\[0.5ex]
$K, \Gamma$ & $c_s<c$ and $M_{\mathrm{max}}>2~\mathrm{M}_{\odot}$ \\[0.5ex]
\hline
\end{tabular}
\end{center}
\caption{Ranges of input parameters.}
\label{tab:mcmc-paras}
\end{table} 

\subsection{Statistical analysis}
\label{sec:stat}

Motivated by findings from Sec.~\ref{sec:hc} and \ref{sec:dU-Sf}, we explore more parameter space in the attempt to quantify uncertainties in key ingredients discussed above which determine thermal states of transiently-accreting neutron stars. These quantities are listed in Table~\ref{tab:mcmc-paras}, with individual range of values to be applied in the Markov chain Monte Carlo simulations. The likelihood function is constructed from the Aql X-1 data point plus a Fermi function representing the upper limit coming from SAX J1808 (see discussion in Appendix~\ref{app:mcmc} for details).

Our results are summarized in Figs.~\ref{fig:L-dU-SF},~\ref{fig:L-etaQ-eos-mass} and~\ref{fig:Lum-Tcore}, for two selected nuclear matter EoSs (SLy4 and HHJ) combined with modifications in a polytropic form at high density (\Eqn{eqn:eos_para}). The plots show probability distributions for each parameter plane (with arbitrary normalization) and contours corresponding to  $1\,\sigma$ uncertainties. 

\noindent $\bullet$ {\uline{Direct Urca threshold and superfluidity}.} The top panel of Fig.~\ref{fig:L-dU-SF} illustrates favored regions of the unknown direct Urca threshold with some broadening. For both EoSs, the peak probability sits around $\nb^{\rm dU}=0.5\sim0.6~\rm{fm}^{-3}$, although HHJ indicates a lower broadening factor ($\al\lesssim 0.05$) than SLy4 ($\al\approx 0.1$). The bottom panel manifests preference of critical temperatures, $T_{\rm cnt}^{\rm max}$ and $T_{\rm cps}^{\rm max}$, for the $n \sft$ and $p \sfs$ superfluids respectively. Both plots show that the data prefer smaller superfluid gaps, of which the tendency for $n \sft$ is mostly driven by the low temperature observed for SAX J1808 (already obvious in e.g. Fig.~\ref{fig:heating-curves-1}, as $n \sft$ tends to suppress direct Urca neutrino emission that helps explain cold sources); for proton $\sfs$ superfluid Fig.~\ref{fig:L-dU-SF} indicates proton critical temperatures
below $T_{\rm cps}^{\rm max}\approx10^8 K$, much lower than often discussed. In order to understand this feature we gradually raise the value of the proton singlet gap at the best-fit point with other physical inputs fixed, and find that the consequent luminosity of Aql X-1 turns out to be increasing, while that of 1808 remains unaffected. This is evidence of the $p\sfs$ pairing suppressing the dominant cooling process for warmer sources like Aql X-1, modified Urca, and rendering higher predicted surface luminosity; it does not affect cold sources like 1808 because (near the best-fit point in the parameter space) the proton gap ultimately closes before the density reaching the direct Urca onset. We note that this result is not necessarily in agreement with previous studies on the cooling isolated neutron stars, which assume strong proton $\sfs$ superfluid with $T_{\rm cps}^{\rm max}\gtrsim(2-3)\times10^9 K$ and constrain neutron $\sft$ primarily from observations of the NS cooling in Cassiopeia A~\cite{Page:2010aw,Shternin11,Taranto:2015ubs}, or perform a complete data fit to temperatures and ages estimated for all observed isolated NSs and predict even larger proton singlet gaps ($T_{\rm cps}^{\rm max}\gtrsim(8-9)\times10^9 K$)~\cite{Beloin17}.
Theoretically, smaller proton gaps are nevertheless
not excluded: proton-neutron correlations cause the Landau effective mass of the proton to be smaller than that of the neutron, and medium polarization effects are much more difficult to take into account for the proton pairing, due to the fact that protons are immersed within the dense neutron background~\cite{Page04}; both effects are expected to reduce the size of the $p\,\sfs$ gap. Beyond the BCS approximation, the maximum value of the gap remains uncertain~\cite{Dean:2002zx,Baldo:2007jx}. Also, we emphasize that anomalous axial contributions to the neutron triplet pairing that PBF neutrino emissivity being suppressed by a factor of $\sim0.19$~\cite{Leinson:2009nu,Leinson:2016dat} have not yet been incorporated in the present paper, which should raise our predicted values of $T_{\rm cnt}^{\rm max}$ (as well as the indicated $n \sft$ gap magnitude) to some extent; a full exploration of this effect will be completed in a follow-up work.

\noindent $\bullet$ {\uline{High-density EoS, crust and envelope}.} The first four plots in Fig.~\ref{fig:L-etaQ-eos-mass} are related to the neutron star composition and structure. The light-element layer thickness of the hot star in Aql X-1, characterized by the value of $\eta$, is centered around $10^{-11}$. This result is not surprising as a layer of light elements helps make Aql X-1 warm without affecting SAX J1808 (we assume no light elements in its envelope). The deep crustal heating power is taken to be the same in both objects; therefore the cold temperature of SAX J1808 prefers smaller values of $Q$ ($1.1\sim1.2 \,\rm{MeV}$ for HHJ EoS and $\sim1.15 \,\rm{MeV}$ for SLy4). We vary the direct Urca threshold
separately from the variation in the EoS (see discussion in Appendix~\ref{app:mcmc}), and find that the data suggest negative values of $K$ and hence a softer EoS at energy density $\gtrsim2\,\ep_0$. To help explain this behavior we increase $K$ from the negative value at the best-fit point until a positive value of roughly the same magnitude, keeping other quantities unchanged, and see how this affects the output of luminosities. Along with this variation we find successive growth in $L_{1808}$ and reduction in $L_{\rm Aql}$, as the central densities are lowered due to stiffening in the EoS at fixed mass, shrinking the regions where direct Urca is operative for massive stars (1808 less colder) and enlarging the radii for low-mass and intermediate-mass stars to induce general relativistic effects similar to those from smaller compactness (Aql X-1 less warmer, see also discussion in Sec.~\ref{sec:hc}). In nucleons-only models without a strong phase transition, a lower direct Urca threshold is correlated with a larger value for the derivative of nuclear symmetry energy (see e.g. Table~\ref{tab:EoS}), and a larger neutron star radius. Our results suggest an opposite trend: both softening of the EoS which implies a small radius~\cite{Lattimer:2000nx} and a not-too-high direct Urca threshold are more desirable simultaneously. In this regard our results are mildly suggestive of exotic matter which can lower the pressure with a smaller direct Urca 
threshold. A complete exploration of cooling models with exotic matter is in progress.

\noindent $\bullet$ {\uline{Mass and core temperature}.} The stringent
constraint provided by the observation of 1808 favors direct Urca
process operating in the interior of a massive ($\gtrsim1.6\,\Msolar$) neutron star; on the other hand, recent analysis of pulse shapes detected during multiple outbursts of 1808 indicates best-fit models with
$0.8\,\Msolar<M<1.7\,\Msolar$~\cite{Morsink:2009wv}, and also optical
observations of the radial velocity of the donor star and of the light
curve constraining the inclination indicate a relatively low mass for
1808~\cite{elebert2009optical,wang2013multiband}. If the observational
constraint continues to push the mass of neutron star in SAX J1808 to
smaller values, then the disagreement will demand a new explanation.
The most likely mass for the neutron star in Aql X-1 is close to the
canonical value of $1.4~\mathrm{M}_{\odot}$, which is not yet
constrained by observation~\cite{sanchez2016donor}. For the SLy4
model, the posteriors in the luminosity for Aql X-1 have larger uncertainty, corresponding to the fact that this star is firmly within the neutrino-dominated cooling regime and variations in the nature of the core have a larger impact on the luminosity. For SLy4, the posterior distribution of the luminosity of SAX J1808 is strongly peaked at low luminosities as we expect, and is strongly correlated with the core temperature as expected from a source where the emission is photon-dominated. For the HHJ model, there is a significant part of the posterior distribution for the luminosity of SAX J1808 which lies at relatively large luminosities. This result is representative of the fact that the cold surface temperature of SAX J1808 is difficult to fit in this model, even with our large parameter space: there are so many configurations in the parameter space which have a large luminosity for SAX J1808 that they form a significant part of the
posterior density in spite of the exponential suppression implied by
our likelihood (Eq.~\ref{eq:like} in the Appendix~\ref{app:mcmc}).

\noindent $\bullet$ {\uline{Quality of the fit}.} The maximum values of our likelihood function which we obtain are 0.46 for SLy4 and 0.62 for HHJ. This corresponds to a perfect fit for Aql X-1 and a luminosity for SAX J1808 close to the observational upper limit. There is some numerical uncertainty in these maximum likelihoods due to the limited size of the Markov chain, but we do not expect likelihoods equal to unity. From either the frequentist or Bayesian viewpoints, with 14 parameters and only two data points, one might expect a better fit to be found easily. This result comes partially from the fact that many of our parameters (like $\eta$ and $Q$) vary only over a small range because they are constrained from other physical considerations, thus they do not have a strong impact on the fit. However, even with this in mind, the quality of the fit does provide evidence that one of our model assumptions may be incorrect, for example, our assumption that the neutron star is composed only of neutrons, protons and electrons. We plan to address this in future work.

Note that computing the ``chi squared'' per degree of freedom makes no sense in this framework because formally there are more parameters than data points. Including the additional data points in Fig.~\ref{fig:hc-vary-eos-Tc} would not change this fact as each additional star at least requires one new additional parameter to describe its mass. It is difficult to convert this into an over-constrained system without strong assumptions about the nature of superfluidity at high densities or the strong assumptions about the operation of the direct Urca process at densities well beyond those where theoretical uncertainties are under control.

\begin{figure*}[htb]
\parbox{0.5\hsize}{
\centerline{\large SLy4 + Polytropes}
\includegraphics[width=0.9\hsize]{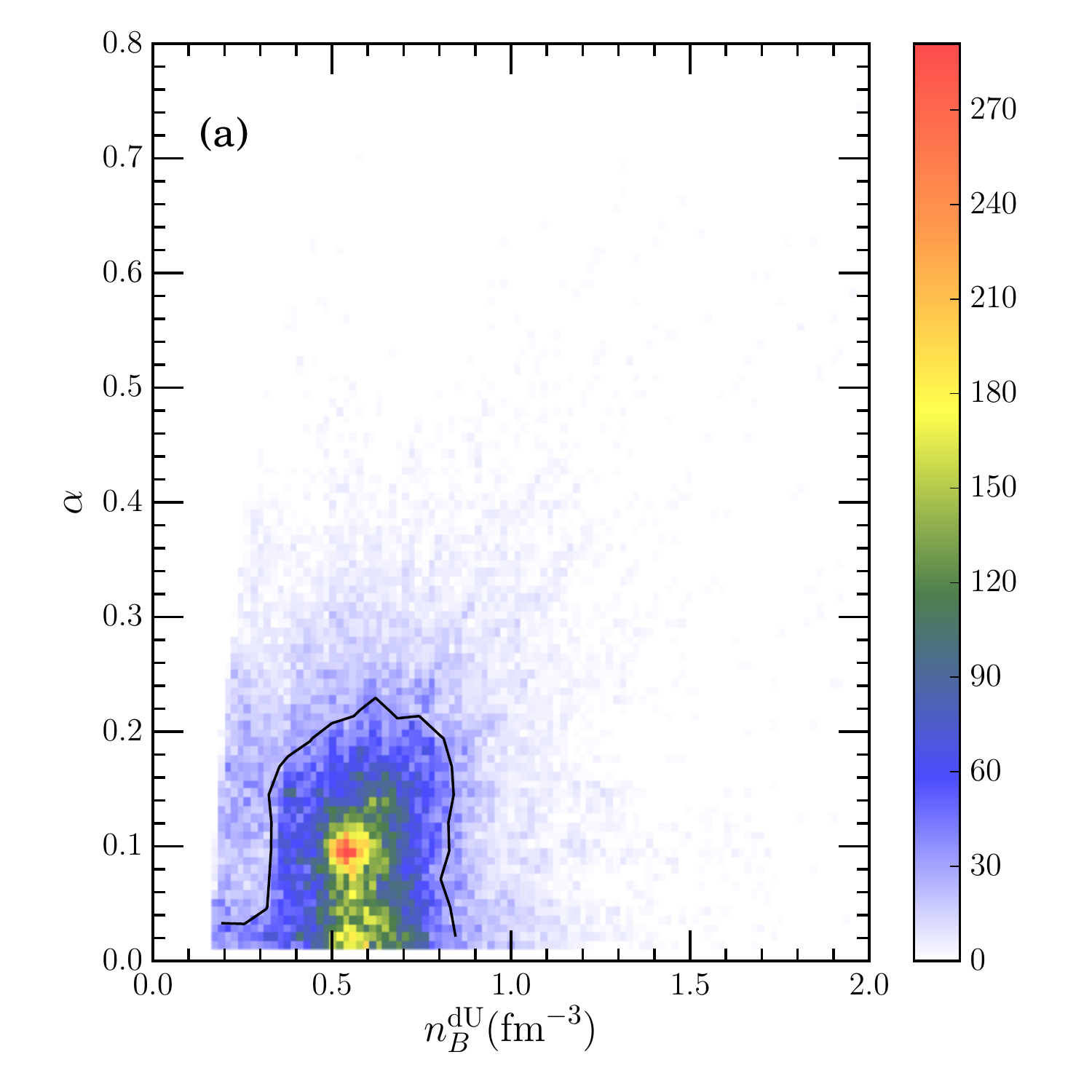} \\[-1ex]
}\parbox{0.5\hsize}{
\centerline{\large HHJ + Polytropes}
\includegraphics[width=0.9\hsize]{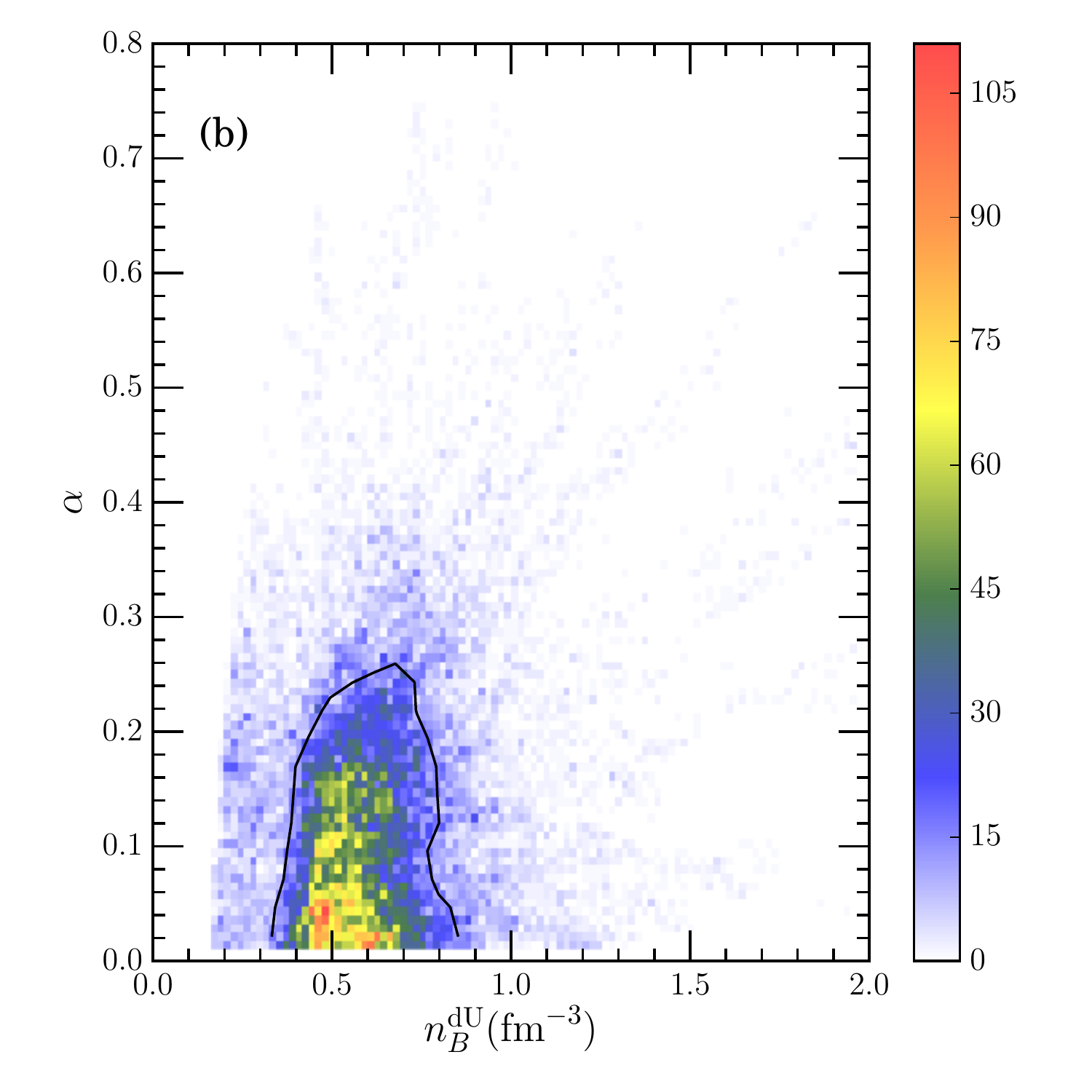} \\[-1ex]
}
\parbox{0.5\hsize}{
\includegraphics[width=0.9\hsize]{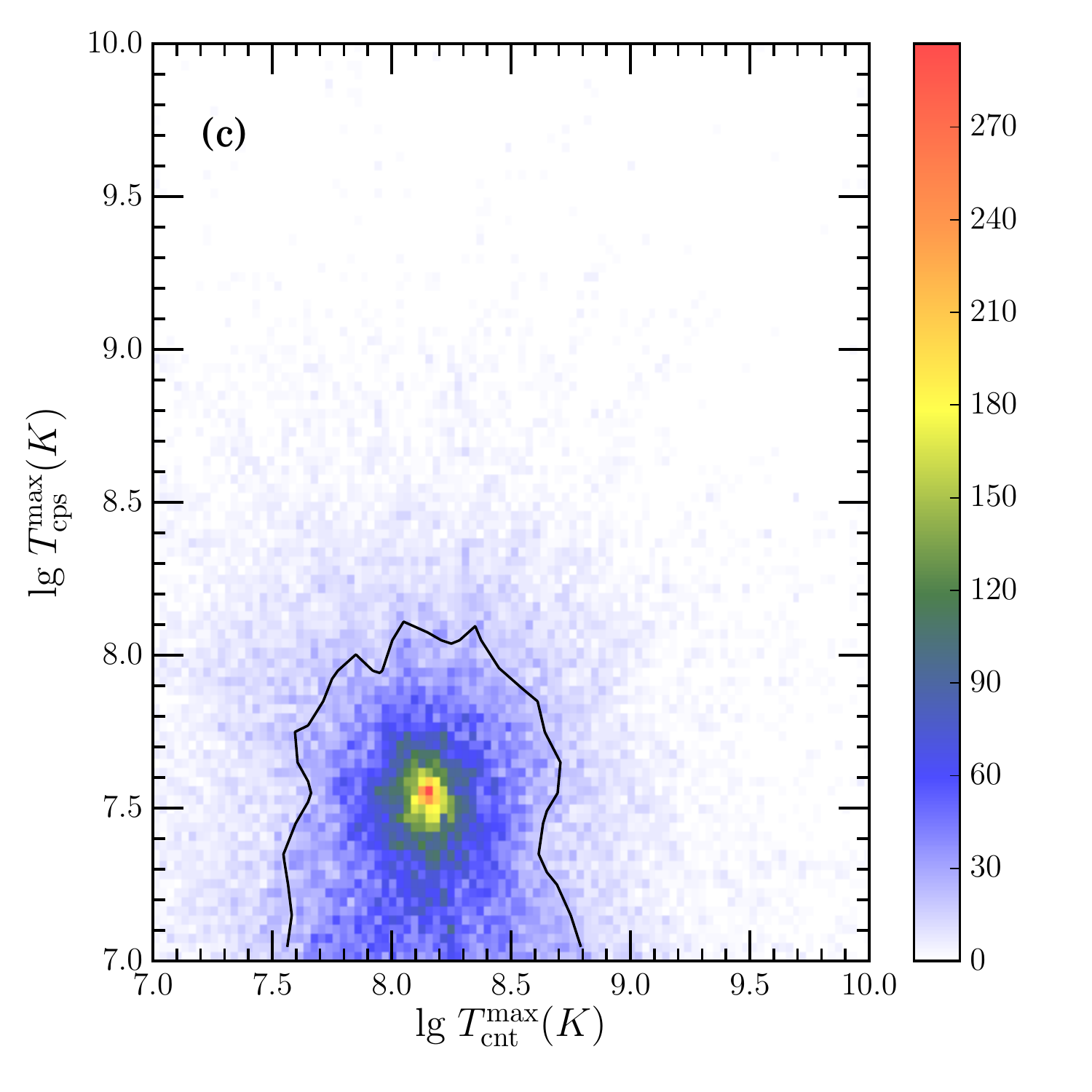} \\[-2ex]
}\parbox{0.5\hsize}{
\includegraphics[width=0.9\hsize]{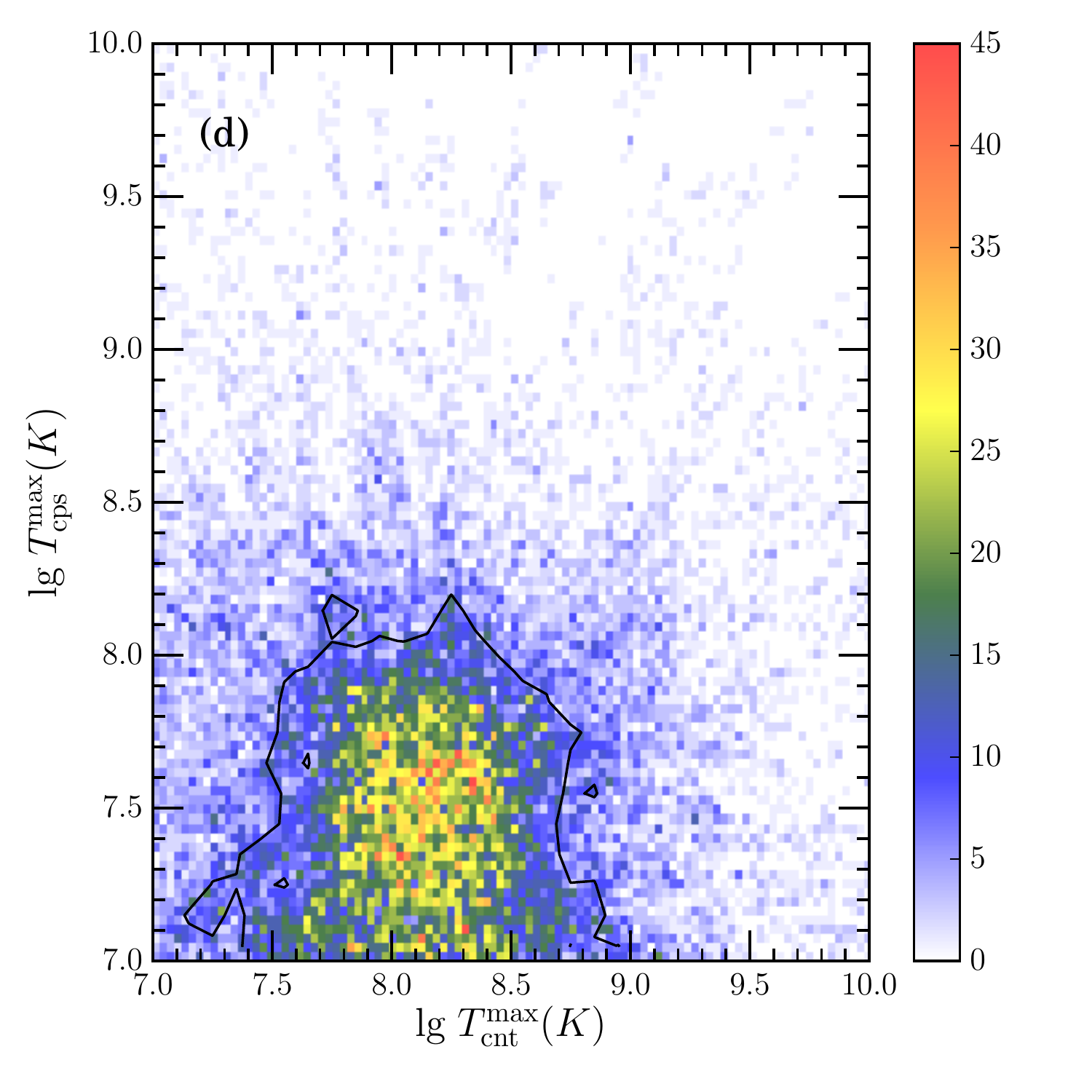} \\[-2ex]
}
\caption{(Color online) Posterior probability distributions for the direct Urca threshold with broadening and $p\sfs$, $n\sft$ superfluid critical temperatures $\mathrm{log}_{10} T_{\rm cps}^{\rm max} - \mathrm{log}_{10} T_{\rm cnt}^{\rm max}$; the thin lines are $1\,\sigma$ contours.
}
\label{fig:L-dU-SF}
\end{figure*}

\begin{figure*}[htb]
\parbox{0.5\hsize}{
\centerline{\large SLy4 + Polytropes} 
\includegraphics[width=0.9\hsize]{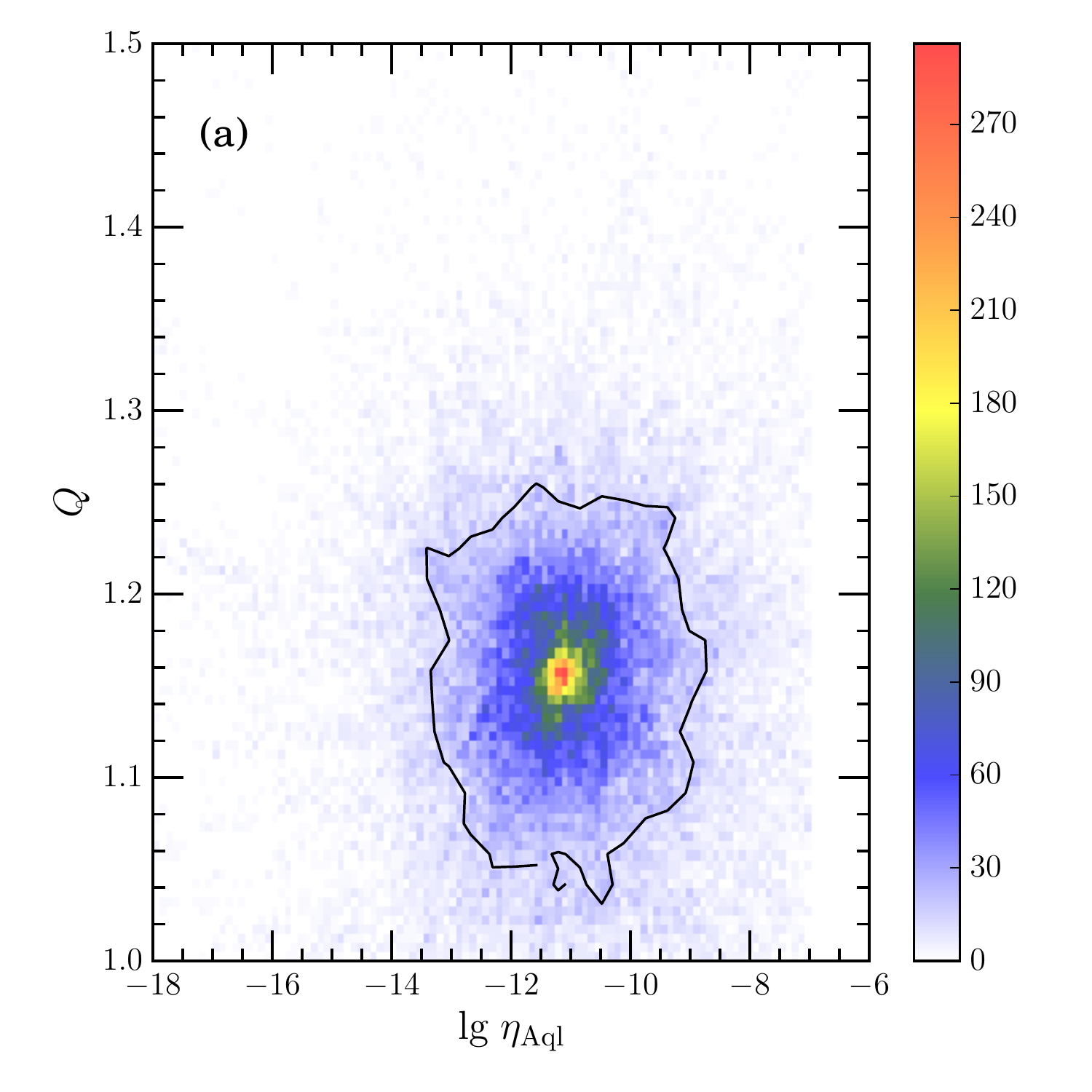} \\[-2ex]
}\parbox{0.5\hsize}{
\centerline{\large HHJ + Polytropes} 
\includegraphics[width=0.9\hsize]{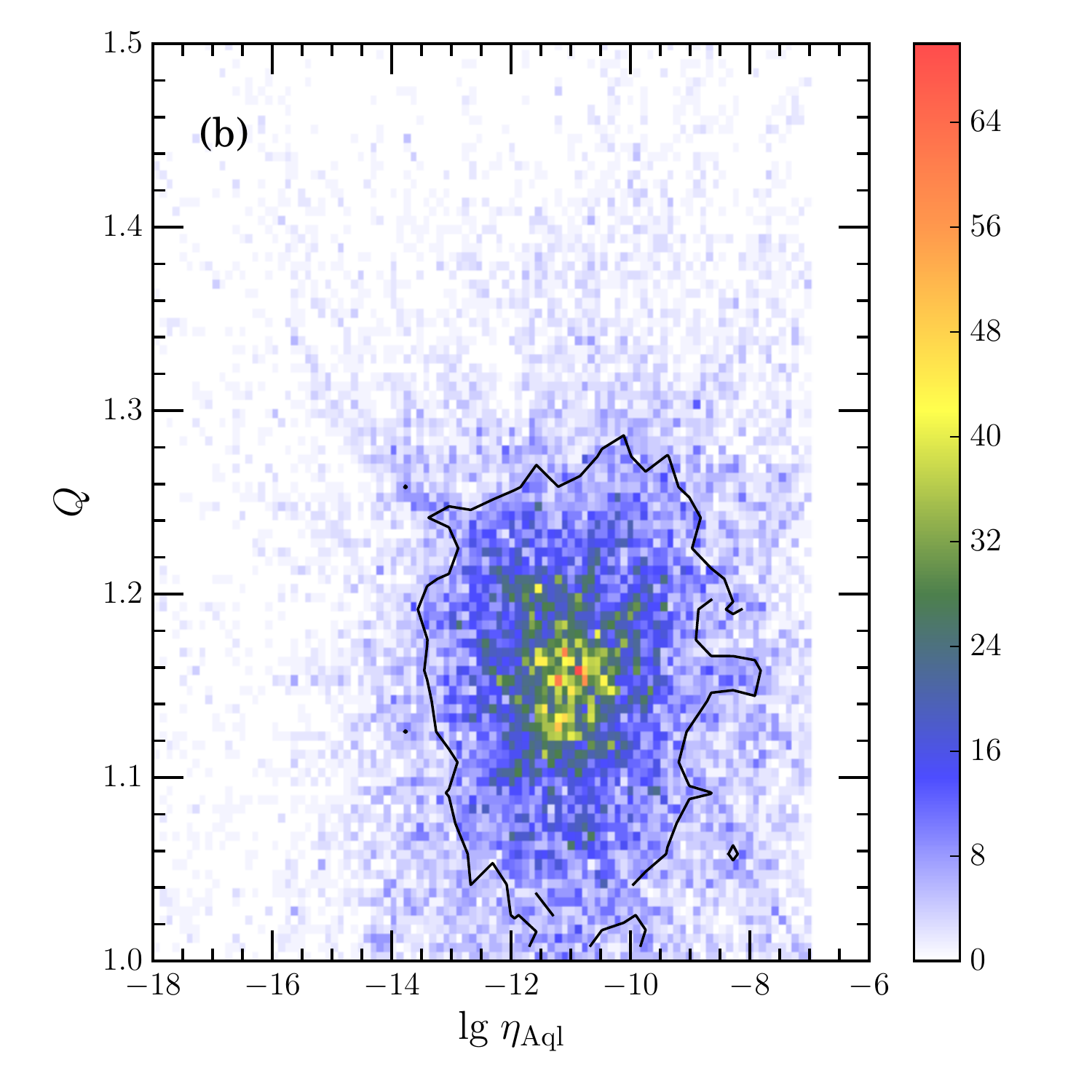} \\[-2ex]
}
\parbox{0.5\hsize}{
\includegraphics[width=0.9\hsize]{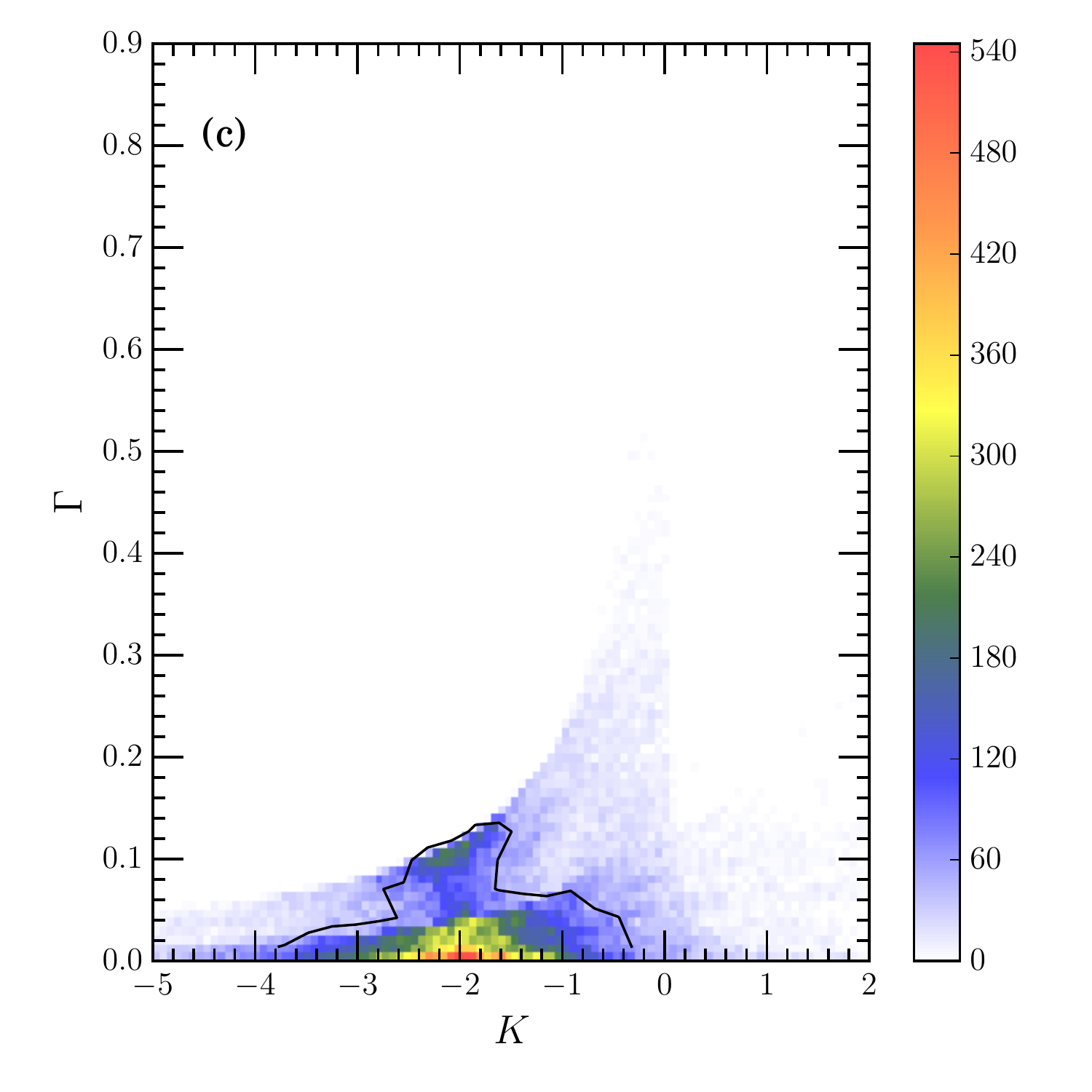} \\[-2ex]
}\parbox{0.5\hsize}{
\includegraphics[width=0.9\hsize]{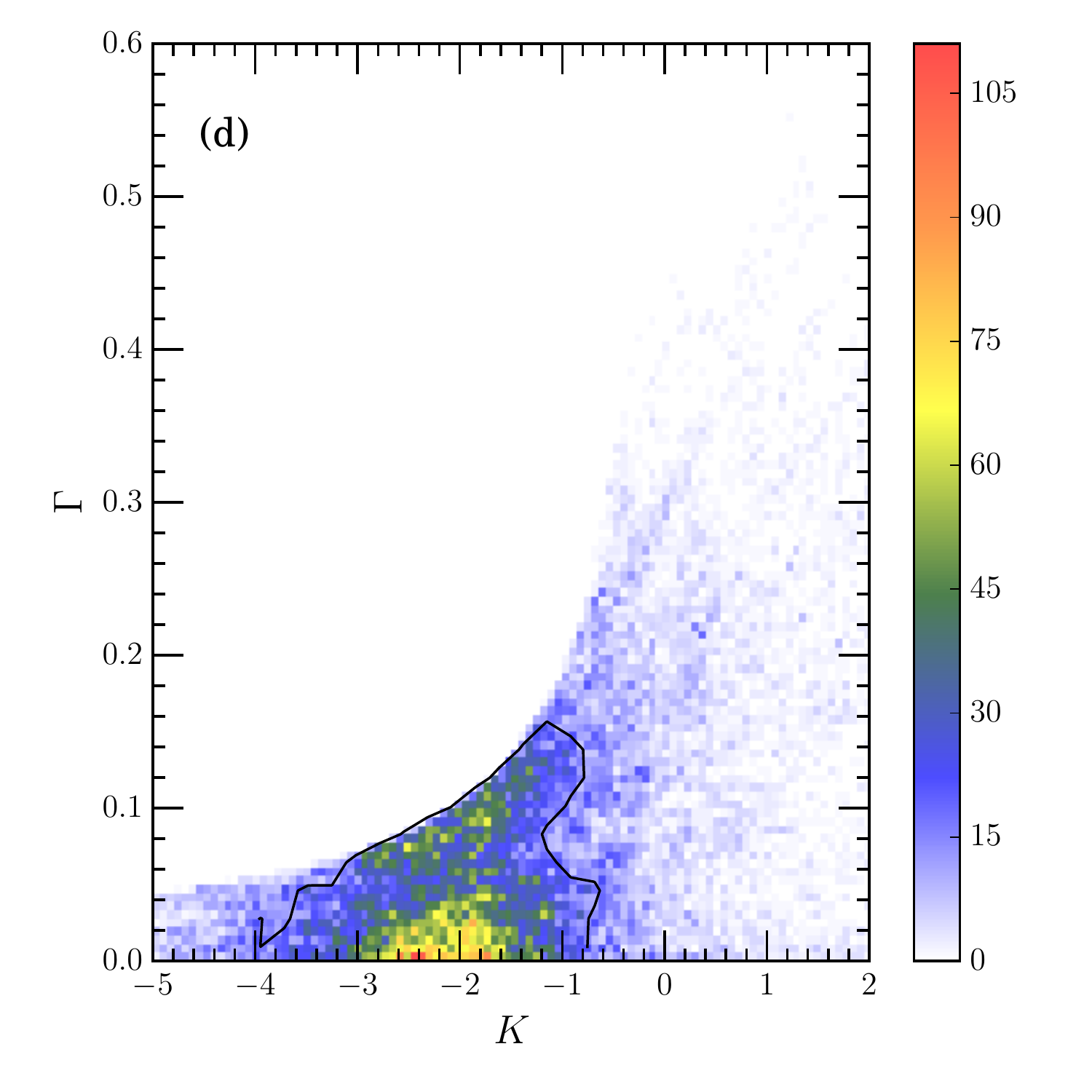} \\[-2ex]
}
\parbox{0.5\hsize}{
\includegraphics[width=0.9\hsize]{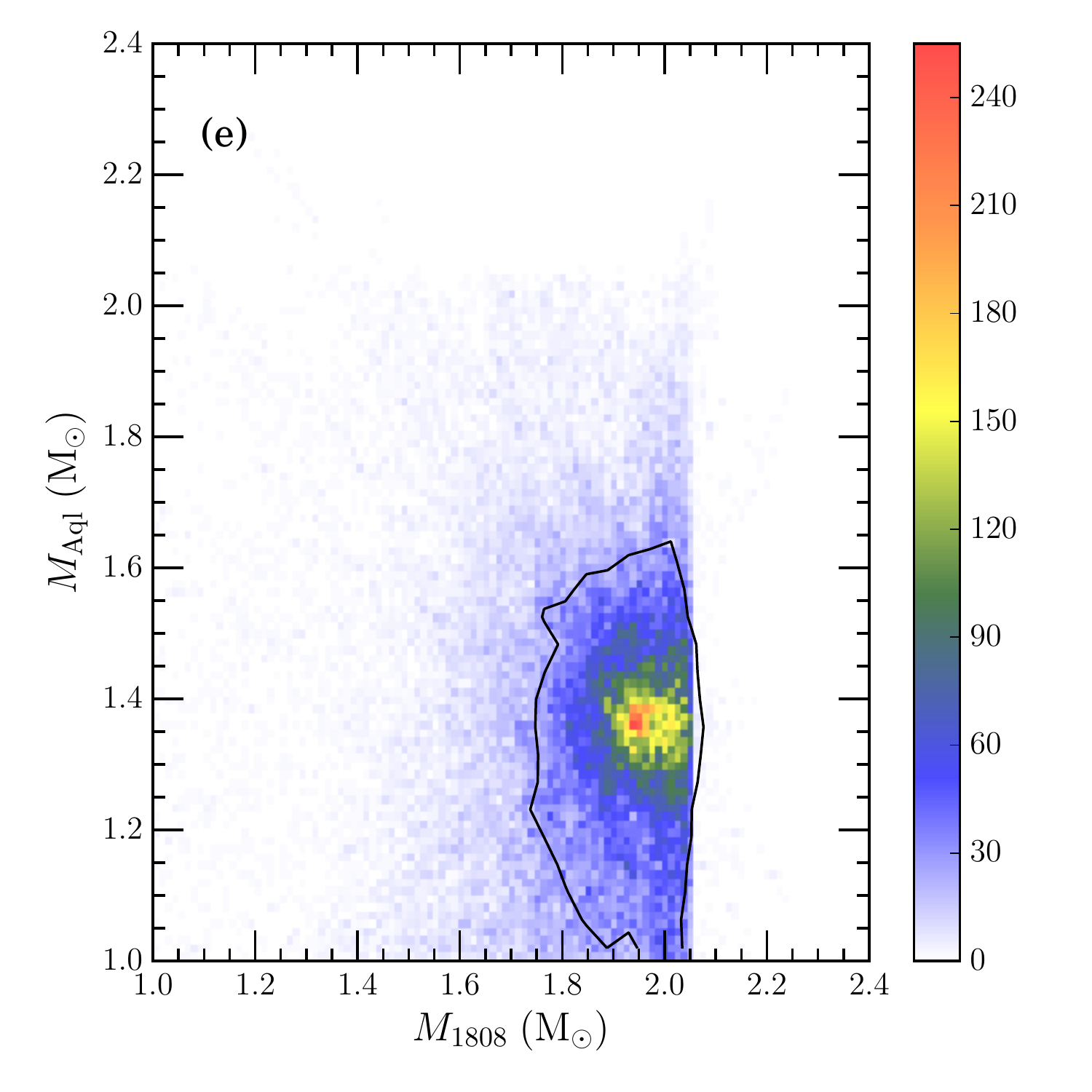} \\[-2ex]
}\parbox{0.5\hsize}{
\includegraphics[width=0.9\hsize]{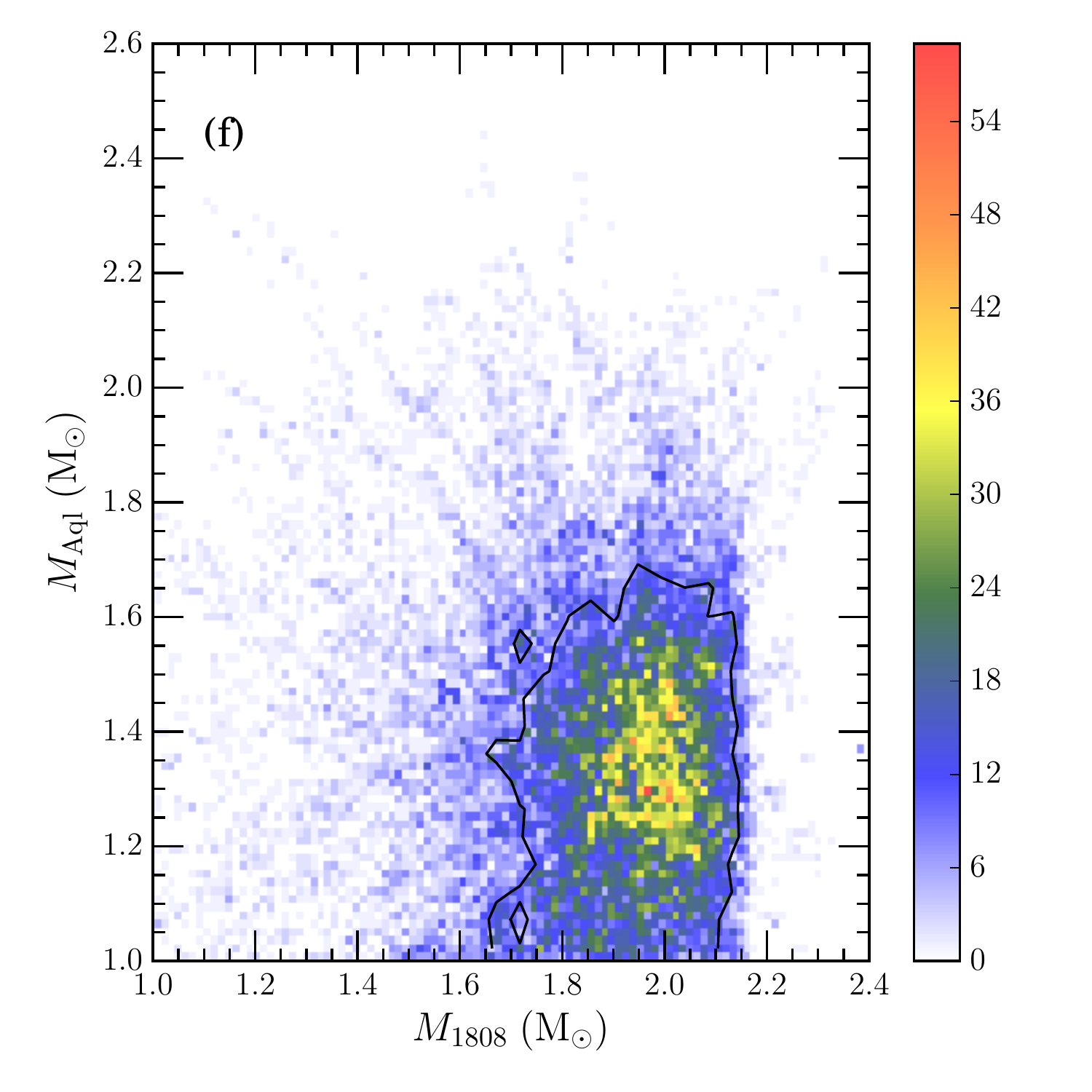} \\[-2ex]
}
\caption{(Color online) Posterior probability distributions for light-element amount $\mathrm{log}_{10} \eta_{\rm Aql}$, deep crustal heating energy $Q$, high-density EoS and masses of neutron stars in 1808 and Aql X-1; the thin lines are $1\,\sigma$ contours.
}
\label{fig:L-etaQ-eos-mass}
\end{figure*}

\begin{figure*}[htb]
\parbox{0.5\hsize}{
\centerline{\large SLy4 + Polytropes}
\includegraphics[width=0.95\hsize]{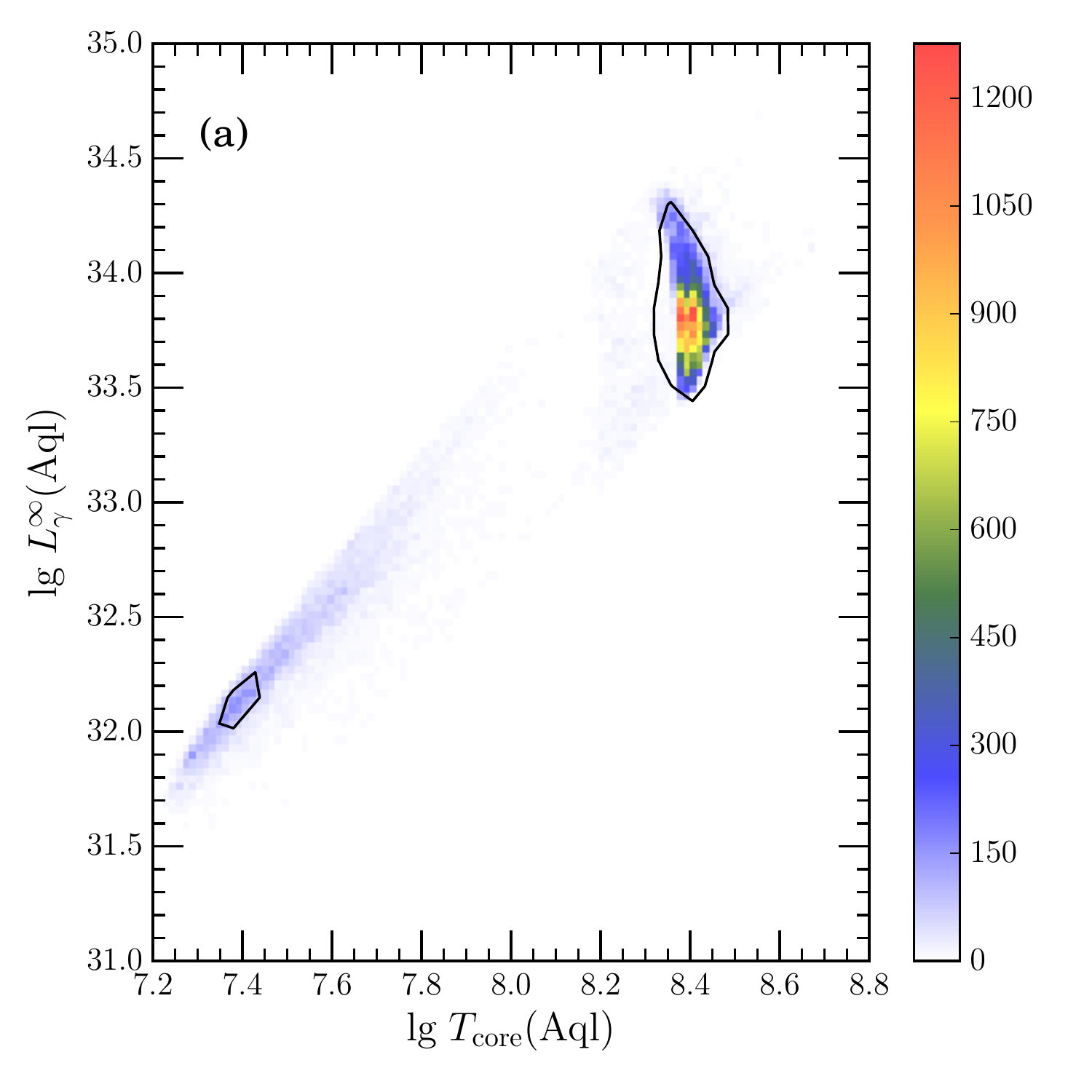} \\[-1ex]
}\parbox{0.5\hsize}{
\centerline{\large HHJ + Polytropes}
\includegraphics[width=0.95\hsize]{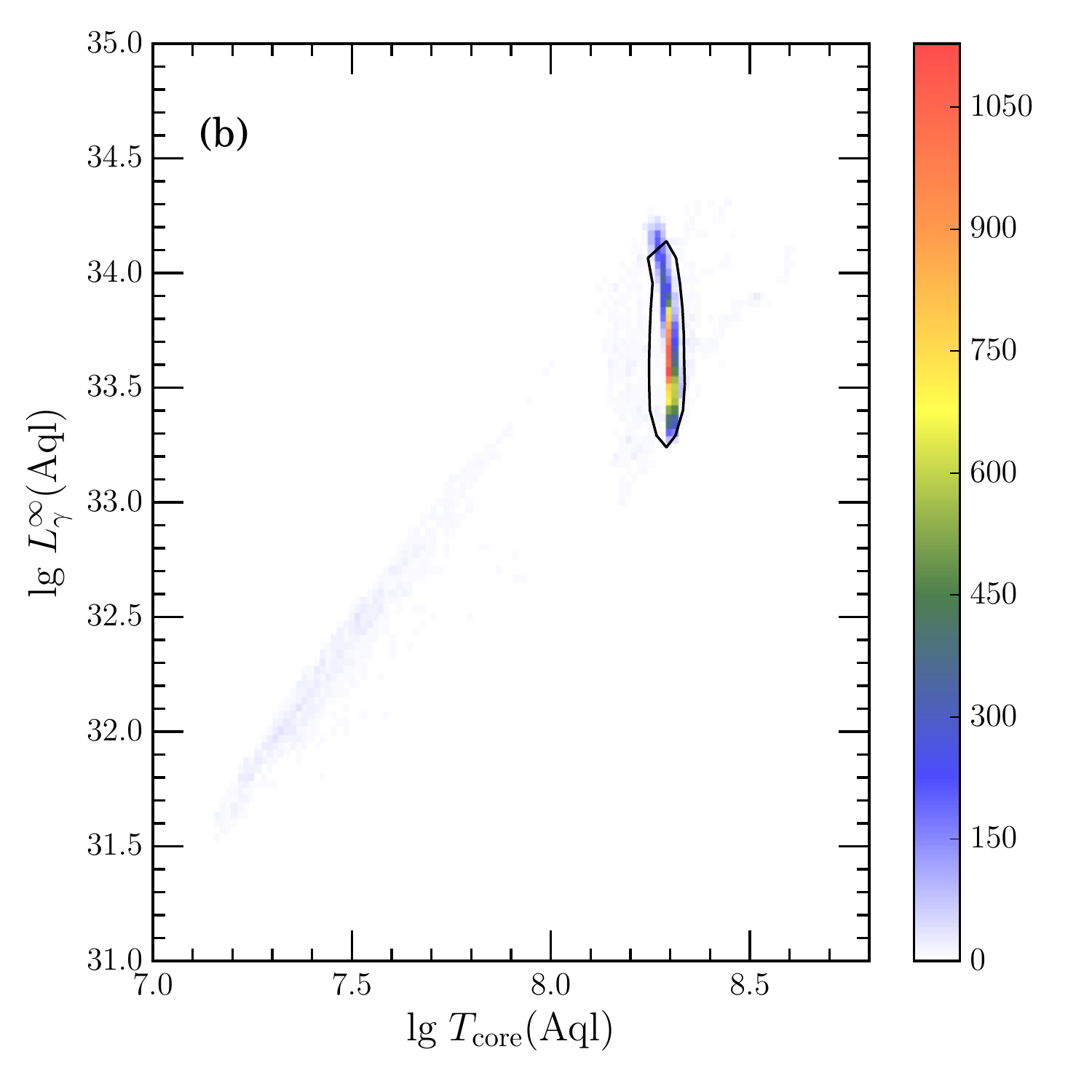} \\[-1ex]
}
\parbox{0.5\hsize}{
\includegraphics[width=0.95\hsize]{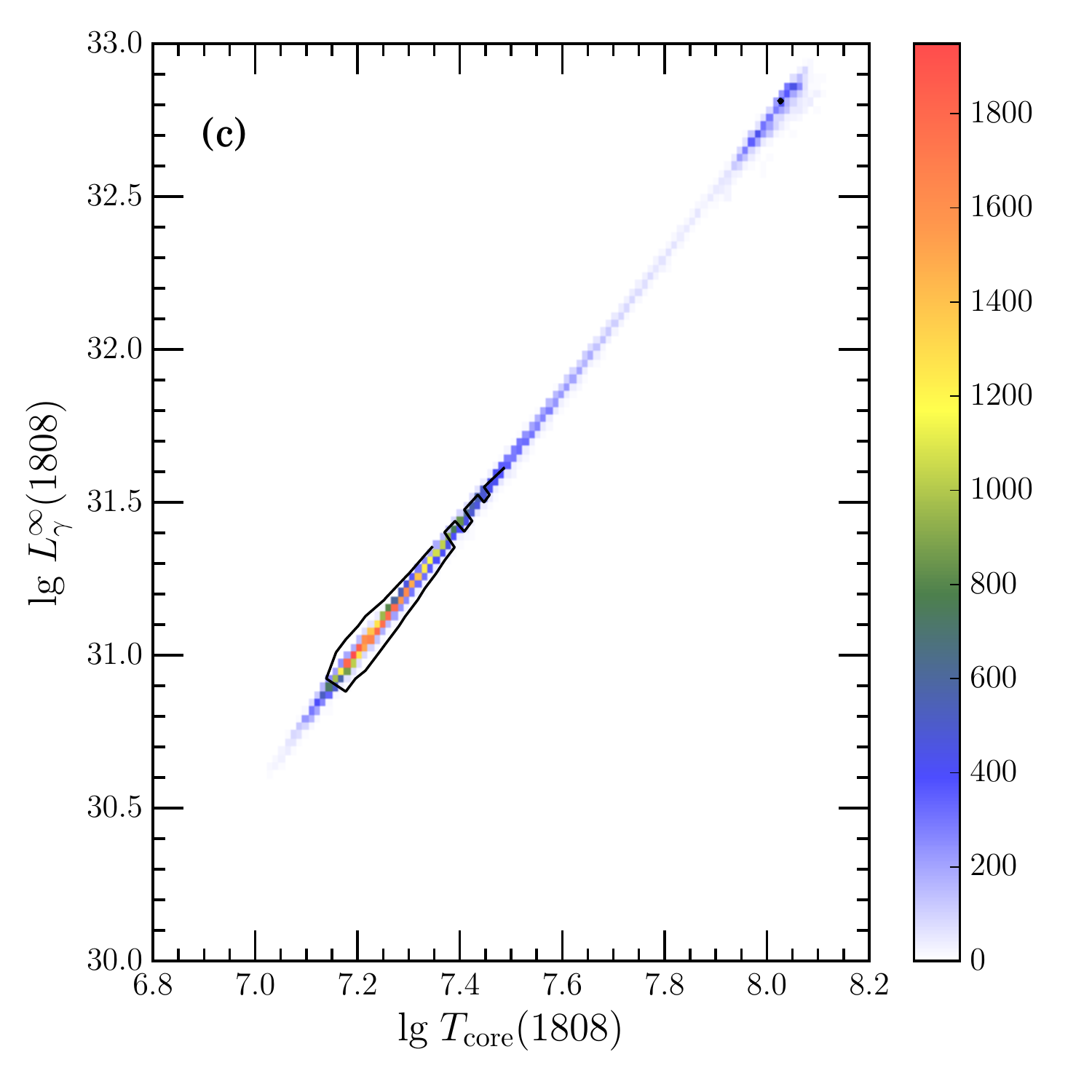} \\[-2ex]
}\parbox{0.5\hsize}{
\includegraphics[width=0.95\hsize]{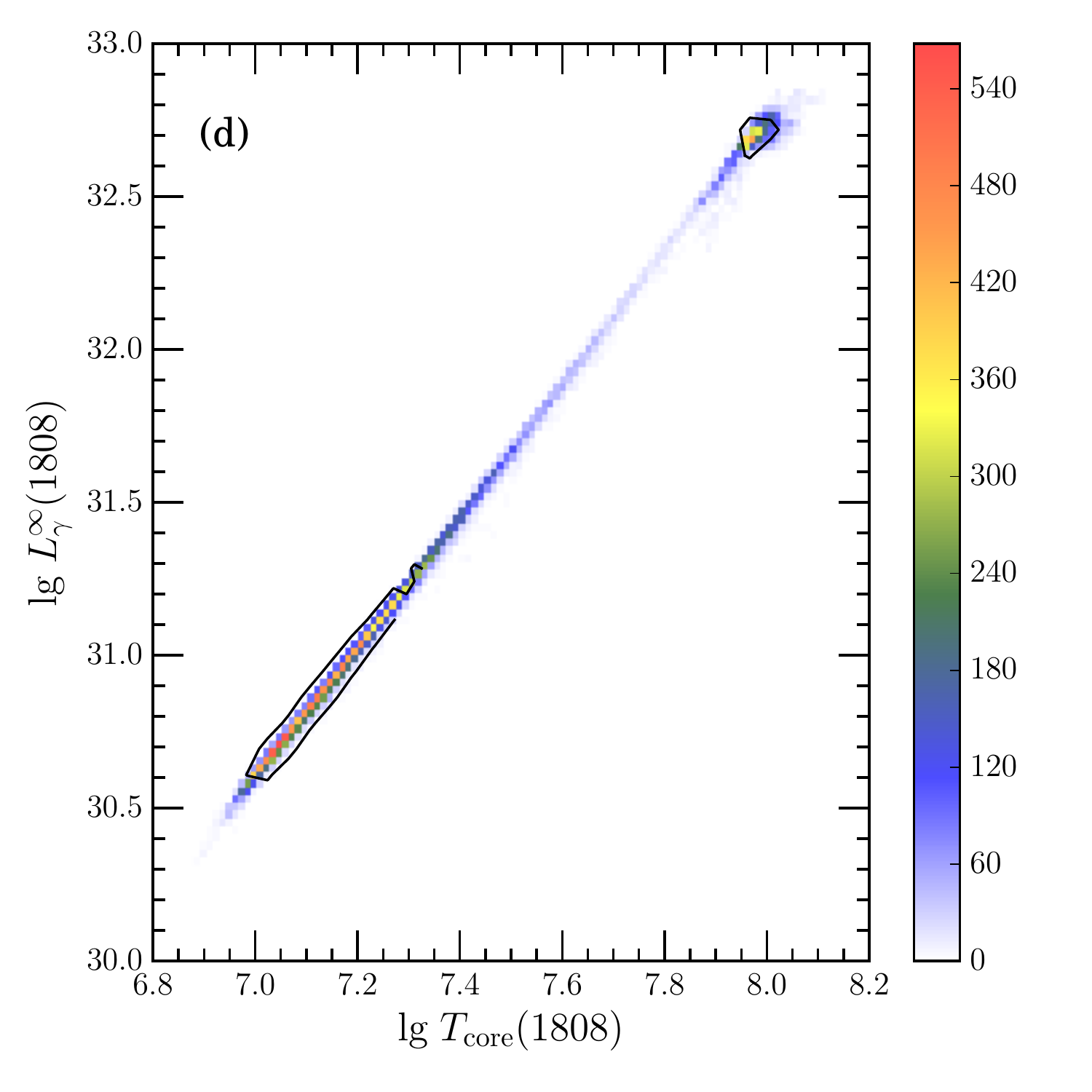} \\[-2ex]
}
\caption{(Color online) Posterior probability distributions for the surface luminosity $\mathrm{log}_{10} L_{\ga}^{\infty}$ and core temperature $\mathrm{log}_{10} T_{\rm core}$ of neutron stars in 1808 and Aql X-1; the thin lines are $1\,\sigma$ contours.
}
\label{fig:Lum-Tcore}
\end{figure*}

\section{Conclusions}
\label{sec:con}

In this paper we have studied thermal states of transiently accreting neutron stars in the quasi-stationary scenario (cooling primarily through neutrino emission balanced by deep crustal heating in the accreted matter), depicted by the heating curves on the $L_{\ga}^{\infty}-\dot{M}$ diagram. By comparing these curves to available observational data of SXRTs, we elucidated the statistical approach to quantify uncertainties in theoretical parameter values.

We have restricted our calculations to the regime of purely-hadronic matter. We take representative nuclear matter EoSs, with typical gap models describing neutron $\sfs$, neutron $\sft$ and proton $\sfs$ superfluids in the star. We also consider the presence of light-element (H or He) layer on top of the accreted iron crust, with its maximum amount being limited by pycnonuclear transformations into heavier elements.

With these models, we present surface luminosity predictions for accreting stars in a quiescent state (Figs.~\ref{fig:hc-vary-eos-Tc}, \ref{fig:heating-curves-1} and~\ref{fig:heating-curves-eta}) at its time-averaged accretion rate. The explanation of two most important sources that impose tight constraints, Aql X-1 and SAX J1808.4-3658, implies that relatively small 
(neutron triplet) superfluid gaps and direct Urca process operating in most massive stars are preferable. Observations of warmer stars can provide a handy test of the deep crustal heating argument, estimate the light element residue in the heat-blanketing envelope, and help discern the EoS at densities pertinent to low-mass and intermediate-mass neutron stars; those of very cold stars reflect the properties of superfluid at the densest cores, and the fastest cooling mechanism, direct Urca process. In the latter case, we also considered a possible broadening of the direct Urca threshold, which turns out to alleviate a bit more the tight constraint from e.g. 1H 1905+000 and SAX J1808.4-3658 (Figs.~\ref{fig:R_dU},~\ref{fig:emissivity-profiles} and~\ref{fig:heating-curves-2}). These results are in fairly good agreement with previous studies~\cite{Beznogov:2014yia,Beznogov:2015ewa,Heinke:2008vj,Yakovlev:2002ti,Yakovlev:2003ed}.

Following this method, we proceed with categorization in all physical inputs 
(Table~\ref{tab:mcmc-paras}) and finally demonstrate statistical analyses on the probability distribution among these parameters (Figs.~\ref{fig:L-dU-SF},~\ref{fig:L-etaQ-eos-mass} and~\ref{fig:Lum-Tcore}). For the two cases 
investigated, we find that (i) direct Urca is likely switched on around $\nb^{\rm dU}(1-\alpha)\sim3\,n_{0}$, and the $p\sfs$ and $n\sft$ superfluids in the core region tend to have small critical temperatures ($T_{\rm cps}^{\rm max}\lesssim10^8 K$ and $T_{\rm cnt}^{\rm max}\lesssim10^9 K$). Small neutron triplet gaps ensure that quite massive stars undergo adequately fast neutrino cooling (thus low surface luminosity), whereas small proton singlet gaps prevent lighter-mass stars from being even brighter than Aql X-1. These predicted values can be adjusted if more cold and/or hot accreting NSs in SXRTs were to be detected in the future; (ii) energy release from deep crustal heating $Q=1\sim1.3 \,\rm{MeV}$ and finite amount of light elements 
$\eta=10^{-13}\sim 10^{-8}$ are able to produce high luminosity observed in Aql X-1; (iii) high-density ($\nb>2\,n_{0}$) polytropic EoS exhibits a softening compared to the original nuclear EoS; and (iv) masses and core temperatures vary vastly for hot and cold stars as expected. These findings are a first systematic assessment of various input physics in this steady-state context, giving valuable insights into the future application of statistical tools.

The data shown in Fig.~\ref{fig:hc-vary-eos-Tc} cover a wide range of luminosities in between theoretical heating curves, which have a gap (see e.g. Fig.~\ref{fig:heating-curves-2}) between stars that are massive enough to cool through the direct Urca process and those less massive stars which do not. This implies that a significant amount of fine-tuning of the neutron star mass may be required to reproduce data 
between these two extremes. The best way to quantify this fine-tuning is to compare two models (for example by computing the relative Bayes factor between them) and show that one model requires much more fine-tuning than the other. It is possible that the appearance of hyperons will broaden the mass range for transitions between slow and fast cooling, but this is beyond the scope of the present paper.

Recently Ref.~\cite{Ofengeim:2016rkq} proposed a study on analytical approximation of the heat capacity and neutrino luminosity for both fully non-superfluid matter and the case of strongly superfluid protons with other particles being in normal states, which were then applied to the quiescent states of SXRTs. Their results show that heating curves predominated by direct Urca cooling may achieve even lower luminosities than the coldest source, assuming that direct Urca operates in the entire neutron star core (an overestimation in general). This treatment and its resulting conclusion are in agreement with our analysis, in that the whole core radiating in direct Urca is effectively an even lower limit ($\nb^{\rm dU}\leq \nb^{\rm crust/core}\approx0.08~\rm{fm}^{-3}$) than what we have imposed on the shifting and broadening parameters of the threshold density ($\nb^{\rm dU}(1-\alpha) \geq0.16~\rm{fm}^{-3}$ in Table~\ref{tab:mcmc-paras}), therefore not surprisingly suffices to produce lower luminosities than our calculations. Ref.~\cite{Matsuo:2016sro} investigated 
enhanced cooling through strong pion condensates in neutron stars, but effects of superfluidity were neglected. In this study they found that SAX J1808.4-3658 was associated with a much lighter mass ($0.75-1.28\,\Msolar$), in striking contrast to the values predicted by purely-hadronic models.

Finally, let us add the caveat that the current work is still under several simplifications and can be improved in the following aspects. On the one hand, the direct Urca onset and its neutrino emissivity are artificially moderated, which ought to be handled properly in a consistent and more physical manner; similarly, many-body corrections to the emissivity of modified Urca~\cite{DehghanNiri:2016cqm} are to be incorporated in the simulations. On the other hand, superfluid gaps in a generic Gaussian form are completely phenomenological, ignoring sophisticated adjustments necessary for some realistic models. For instance, it has been shown that cooling rate can have remarkable dependence on the multicomponent 
phase state of $\sft$ superfluid neutrons~\cite{Leinson:2014cja}. In addition, as mentioned earlier, exotic matter may lead to exceedingly different interpretation of thermal states if they are present in neutron stars. We hope to extend the scope of our work to include kaon/pion condensates, hyperons and strange quark matter in the future.

\section*{Acknowledgements} 

The authors gratefully thank Craig Heinke for valuable discussions and comments on an early version of the manuscript, and acknowledge the anonymous referee for insightful suggestions and advices that significantly improved this paper. S.H. acknowledges the program ``The Phases of Dense Matter'' (INT-16-2b) held at Institute for Nuclear Theory, University of Washington, Seattle where part of this work was completed. S.H. and A.W.S. were supported by grant NSF PHY 1554876 and the U.S. DOE Office of Nuclear Physics. This project used computational resources from the University of Tennessee and Oak Ridge National Laboratory's Joint Institute for Computational Sciences.

% macros used by ADS Database BiBTeX entries:
% see http://adsabs.harvard.edu/abs_doc/aas_macros.sty
\newcommand{\apjl}{Astrophys. J. Lett.\ }
\newcommand{\mnras}{Mon. Not. R. Astron. Soc.\ }
\newcommand{\aap}{Astron. Astrophys.\ }

\bibliographystyle{apsrev}
\bibliography{sxrt}

\appendix
\section{MCMC simulations (SLy4/HHJ EoS)}
\label{app:mcmc}

In order to characterize the nature of models which are able to explain the observational data and to obtain parameter uncertainties, we perform a Markov chain Monte Carlo simulation based on the likelihood function from which fits SAX J1808 and Aql X-1. In principle, the associated likelihood functions ought to be constructed according to the photon counting statistics from which the upper limit and/or mean value are derived; however, due to the lack of full probability distribution functions it is difficult to determine an ideal fit that can integrate specific measurements of individual sources presumably. Because the observational uncertainties are often given as multiplicative factors and the luminosities are positive, here we adopt the logarithms of the luminosities in the likelihood
\begin{eqnarray}
  {\cal L} &=& e^{-\left[\mathrm{log}_{10} L(\dot{M}_{\mathrm{Aql}})-
      \mathrm{log}_{10} L_{\mathrm{Aql}}\right]^2/2/
    \delta L_{\mathrm{Aql}}^2} \nonumber \\
  && \times \left\{ 1 +
  e^{\left[\mathrm{log}_{10} L(\dot{M}_{1808})-
      \mathrm{log}_{10} L_{\mathrm{1808}}\right]/
    \delta L_{\mathrm{1808}}}\right\}^{-1}
  \label{eq:like} \;\;
\end{eqnarray}
where $\dot{M}_{\mathrm{Aql}}=4 \times 10^{-10}\,\Msolar/\mathrm{yr}$,
$\dot{M}_{\mathrm{1808}}=10^{-11}\,\Msolar/\mathrm{yr}$, $\delta
L_{\mathrm{Aql}} = 0.5$ (half a decade uncertainty) and we choose
$\delta L_{\mathrm{1808}} = 0.5$ as well. This is a Gaussian data
point in log-log space for the luminosity of Aql X-1 and an upper limit for the log of the SAX J1808 luminosity represented by a Fermi distribution function. The normalization is such that if the luminosity for Aql X-1 is matched exactly and the luminosity for SAX J1808 is equal to the upper limit, the likelihood is 1/2. If the luminosity for SAX J1808 is much smaller than the upper limit, the likelihood takes a maximum value of unity. Our choice of the fit function in the logarithmic space is not unique, and can be improved with better-known statistics from the observation in the future. Following Ref.~\cite{Beloin17}, we describe the gap using 3 parameters for the proton singlet gap and 3 parameters for the neutron triplet gap.
Additional parameters are the two neutron star masses, the deep crustal heating parameter $Q$, a parameter $\eta$ specifying the amount of light elements in the envelope of Aql X-1, the threshold density for the direct Urca process, and the broadening parameter for direct Urca. Finally, the equations of state which we use near the nuclear saturation density may not be correct at higher densities. To modify the EoS at high densities, we use the following parametrized form 
\beq 
P=P_{\rm{NM}} (\ep) + \Theta (\ep - 2\ep_0)K\left[\ep^{\Ga}-(2\ep_0)^{\Ga}\right],
\label{eqn:eos_para}
\eeq
where $\ep_{0}=\ep_{\rm NM}(n_{\rm B} = n_0)$ is the energy density in
the nuclear matter EoS at saturation. This form contains two new parameters, $K$ and $\Gamma$ for a total of 14 parameters. 

For each Monte Carlo point, we must ensure that several restrictions are satisfied. We ensure that the pressure is non-decreasing at all densities and that the EoS is causal, i.e. the sound speed is less than the speed of light at all densities below the central density of the maximum-mass star. In accordance with recent observations from Refs.~\cite{Demorest10,
Antoniadis13}, we ensure that the maximum mass is larger than $2\,\Msolar$. In order to avoid double-counting of the superfluid parameters, we ensure that the Fermi momenta for which the critical temperatures are
maximized are above the crust-core transition and below the central
density of the maximum-mass neutron star. Finally, we ensure that the
direct Urca process does not occur at the nuclear saturation density
by imposing $\nb^{\rm dU}(1-\alpha) \geq0.16~\mathrm{fm}^{-3}$. Any MC point that violates any of these conditions is automatically rejected.

We compute posterior probability distributions for the parameters by
marginalization, integrating over the Markov chains which were constructed in the Monte Carlo simulation. Prior probability distributions for all quantities are uniform, except for $\eta$, $T_{\rm cnt}^{\rm max}$, and $T_{\rm cps}^{\rm max}$, for which uniform distributions are used for their logarithms instead. The parameter limits are chosen large enough to significantly affect the posteriors, except for the restrictions given in
Table~\ref{tab:mcmc-paras}. 

We vary the EoS and the threshold density for the direct Urca process
independently, so as to separately vary the pressure and composition
of matter in the neutron star core. In many models, these quantities are not separate but correlated so that an increase in the proton fraction also implies an increase in the pressure. However, this correlation is broken by the possible presence of higher-order terms in the symmetry energy at high
density~\cite{Steiner06hs,Wellenhofer16}, thus we vary them separately.

\end{document}